\documentclass[a4paper,11pt]{article}
\pdfoutput=1 

\usepackage{jheppub} 

\usepackage[T1]{fontenc}

\usepackage{orcidlink}
\usepackage[normalem]{ulem}
\usepackage{color,bm}

\newcommand{\bea} {\begin{eqnarray}}
\newcommand{\eea} {\end{eqnarray}}

\newcommand{\beq} {\begin{equation}}
\newcommand{\eeq} {\end{equation}}

\newcommand{\order}{{\cal O}}

\title{Probing a long-lived pseudoscalar in type-I 2HDM with displaced vertices and jets at the LHC}

\author[a]{Lei Wang\orcidlink{0000-0001-8639-4917}}
\author[b, 1]{Zeren Simon Wang\orcidlink{0000-0002-1483-6314}\note{Corresponding author.}}
\author[a, 2]{Haotian Xu\orcidlink{0009-0006-5131-0385}\note{Corresponding author.}}

\affiliation[a]{Department of Physics, Yantai University, Yantai
264005, China}
\affiliation[b]{School of Physics, Hefei University of Technology, Hefei 230601, China}

\emailAdd{leiwang@ytu.edu.cn}
\emailAdd{wzs@hfut.edu.cn}
\emailAdd{xuhaotian1715@163.com}

\abstract{In the type-I two-Higgs-doublet model, the pseudoscalar $A$ can act as a long-lived particle (LLP) for sufficiently large values of $\tan\beta$. At the LHC, the $A$ particles are predominantly produced in pairs through $pp \to W^*/Z^* \to H^\pm/H \, A$, with subsequent decays $H^{\pm}/H \to W^\pm/Z\, A$. For the mass range of our interest, $10~\text{GeV}\lesssim m_A \lesssim 100~\text{GeV}$, the pseudoscalar $A$ typically decays into a pair of bottom quarks after traveling a macroscopic distance from its production point, giving rise to displaced-vertex (DV) signatures inside the inner detector. We perform Monte Carlo simulations of signal events with DVs plus jets, and assess the discovery prospects of $A$ as an LLP at the ATLAS and CMS experiments. Our findings show that a substantial portion of the parameter space with $m_A>10$ GeV has already been excluded by LHC Run-2 data, while the high-luminosity LHC will be able to probe broader regions.}

\begin{document} 
\maketitle
\flushbottom

\section{Introduction}
\label{sec:intro}

Since the discovery of the Standard-Model (SM) Higgs boson in 2012 at the LHC~\cite{ATLAS:2012yve,CMS:2012qbp}, extensive programs have been conducted to further study its properties which have found results in agreement with the SM expectations; see, for instance, Refs.~\cite{ATLAS:2023dnm,CMS:2022ley,ATLAS:2024wfv,ATLAS:2022vkf,CMS:2022dwd}.
In addition, searching for extra Higgs bosons has become an important task in the quest for physics beyond the SM (BSM). 
Both ATLAS and CMS have carried out comprehensive searches for such new scalar states via their prompt-decay signatures~\cite{ATLAS:2024vxm,ATLAS:2024bzr,ATLAS:2024auw,CMS:2023boe,Jeppe:2024uki,CMS:2024vps}.
However, no clear deviation from SM expectations has been reported.
In recent years, more interest has shifted towards the possibility of new physics manifesting itself as light feebly-interacting particles that are often predicted to be long-lived and therefore have so far escaped the traditional prompt searches at the LHC.
Besides, several far-detector experiments dedicated to searches for such new long-lived particles (LLPs) have been proposed to be operated in the vicinity of various LHC interaction points (IPs), such as FASER and FASER2~\cite{Feng:2017uoz, FASER:2018eoc, FASER:2023tle}, FACET~\cite{Cerci:2021nlb}, CODEX-b~\cite{Gligorov:2017nwh,CODEX-b:2019jve}, MoEDAL-MAPP1 and MAPP2~\cite{Pinfold:2019nqj,Pinfold:2019zwp},  MATHUSLA~\cite{Chou:2016lxi, Curtin:2018mvb,MATHUSLA:2020uve,MATHUSLA:2025zyt,MATHUSLA:2025eth}, and ANUBIS~\cite{Bauer:2019vqk}.

Long-lived scalar particles naturally arise in multiple extensions of the SM, such as the real scalar extension of SM~\cite{Liu:2022nvk}, specific realizations of two-Higgs-doublet models (2HDMs)~\cite{Wang:2022yhm,Kling:2022uzy,Haisch:2023rqs,Liu:2024azc,Shan:2024pcc,Kim:2025tuz,Qi:2025qsj}, Georgi-Machacek model \cite{Lu:2024ade}, neutral-naturalness models~\cite{Burdman:2006tz,Chacko:2005pe,Craig:2015pha,Cai:2008au,Cohen:2018mgv,Cheng:2018gvu}
and non-minimal supersymmetric frameworks~\cite{Ellwanger:2009dp,Cao:2013gba,Adhikary:2022pni}.
In 2HDMs, five mass eigenstates of scalar particles arise from spontaneous electroweak symmetry breaking, including a CP-odd pseudoscalar $A$.
In particular, in type-I 2HDM, the couplings of this pseudoscalar $A$ to SM fermions are suppressed by a factor of $1/\tan\beta$ with $\tan\beta$ being the ratio of the two Higgs doublet's vacuum expectation values (VEVs), making it a viable LLP candidate in the large-$\tan\beta$ regime.
In Ref.~\cite{Kling:2022uzy}, the authors investigated the sensitivity of the FASER and FASER 2 experiments to $A$ produced in meson decays, and found that masses up to $m_A \sim 3$~GeV can be probed.
Also, the authors of Ref.~\cite{Liu:2024azc} explored the potential of several LHC far detectors to probe $A$ via the production processes $pp \rightarrow W^\pm/Z \rightarrow H^{\pm}/H A$, including FASER2, FACET, MAPP2, and MATHUSLA, and found that these experiments can be sensitive to masses up to $m_A \sim 10$ GeV.

In this paper we will focus on the mass range of $10\lesssim m_A \lesssim 100$~GeV in the type-I 2HDM, with the pseudoscalar particle $A$ being long-lived and dominantly decaying into a pair of bottom quarks.
We analyze the discovery prospects of $A$ as an LLP at the ATLAS and CMS experiments, considering both LHC Run-2 data and the upcoming high-luminosity (HL)-LHC, where the long-lived $A$ particles are primarily produced in pairs through $pp \rightarrow W^*/Z^* \to H^{\pm}/H \, A$, followed by $H^{\pm}/H\to  W^{\pm}/Z\,A$.

This paper is organized as follows.
In section~\ref{sec:model}, we provide a brief overview of the type-I 2HDM.
Section~\ref{sec:currentlimits} covers relevant theoretical and experimental bounds on $m_A$ and $\tan\beta$.
Section~\ref{sec:analysis} is devoted to discussing the experimental searches that we recast and propose at the ATLAS and CMS experiments, with the numerical results presented in section~\ref{sec:results}.
Finally, we conclude the work in section~\ref{sec:conclu}.

\section{Type-I two-Higgs-doublet model}
\label{sec:model}

In the 2HDM, the Higgs sector includes two scalar doublets, $\Phi_1$ and $\Phi_2$, both with hypercharge $Y = 1$,
\begin{equation}
\Phi_i=\left(\begin{array}{c} \phi_i^+ \\
\frac{1}{\sqrt{2}}\,(v_i+\phi_i+ia_i)
\end{array}\right).
\end{equation}
Here, $i=1,2$, and $v_1$ and $v_2$ denote, respectively, the VEVs acquired by the two scalar doublets following electroweak symmetry breaking, with the relation $v^2 \equiv v_1^2 + v_2^2 = (246~\rm GeV)^2$.

One may express the CP-conserving Higgs potential under softly broken $Z_2$ symmetry as:
\begin{eqnarray}
\label{eqn:V2HDM}
\mathcal{V}_{\text{tree}} &=& m_{11}^2
(\Phi_1^{\dagger} \Phi_1) + m_{22}^2 (\Phi_2^{\dagger}
\Phi_2) - \left[m_{12}^2 (\Phi_1^{\dagger} \Phi_2 + \rm h.c.)\right]\nonumber \\
&&+ \frac{\lambda_1}{2}  (\Phi_1^{\dagger} \Phi_1)^2 +
\frac{\lambda_2}{2} (\Phi_2^{\dagger} \Phi_2)^2 + \lambda_3
(\Phi_1^{\dagger} \Phi_1)(\Phi_2^{\dagger} \Phi_2) + \lambda_4
(\Phi_1^{\dagger}
\Phi_2)(\Phi_2^{\dagger} \Phi_1) \nonumber \\
&&+ \left[\frac{\lambda_5}{2} (\Phi_1^{\dagger} \Phi_2)^2 + \text{h.c.}\right].
\end{eqnarray}

After spontaneous electroweak symmetry breaking, the physical scalar states include two neutral
CP-even states ($h$ and $H$), one neutral CP-odd pseudoscalar $A$, and a pair of charged scalars $H^\pm$.
The coupling constants in the Higgs potential can be determined from the Higgs mass inputs~\cite{Gunion:2002zf,Kling:2016opi},
\begin{eqnarray}
 &&v^2 \lambda_1  = \frac{m_H^2 c_\alpha^2 + m_h^2 s_\alpha^2 - m_{12}^2 t_\beta}{ c_\beta^2}, \ \ \ 
v^2 \lambda_2 = \frac{m_H^2 s_\alpha^2 + m_h^2 c_\alpha^2 - m_{12}^2 t_\beta^{-1}}{s_\beta^2},  \nonumber \\  
&&v^2 \lambda_3 =  \frac{(m_H^2-m_h^2) s_\alpha c_\alpha + 2 m_{H^{\pm}}^2 s_\beta c_\beta - m_{12}^2}{ s_\beta c_\beta }, \ \ \ 
v^2 \lambda_4 = \frac{(m_A^2-2m_{H^{\pm}}^2) s_\beta c_\beta + m_{12}^2}{ s_\beta c_\beta },  \nonumber \\
 &&v^2 \lambda_5=  \frac{ - m_A^2 s_\beta c_\beta  + m_{12}^2}{ s_\beta c_\beta }\, , 
\label{eqn:lambdas}
\end{eqnarray}
where the shorthand notations are defined as $t_{\beta}\equiv \tan\beta$, $s_{\beta}\equiv \sin\beta$, $c_{\beta} \equiv \cos\beta$, $s_{\alpha}\equiv \sin\alpha$, and $c_{\alpha} \equiv \cos\alpha$ with $\alpha$ being the mixing angle between the two CP-even Higgs bosons and $\tan\beta=v_2/v_1$.
In the limit $\cos(\beta-\alpha) \to 0$, the following approximate relations hold~\cite{Liu:2024azc,Shan:2024pcc,Wang:2021ayg}
\begin{eqnarray}
v^2 \lambda_1 &=&  m_h^2 - \frac{t_\beta^3\,(m_{12}^2 -m_H^2  s_\beta c_\beta ) }{ s_\beta^2}\,,\nonumber \\
v^2 \lambda_2 &=& m_h^2 - \frac{ (m_{12}^2 -m_H^2  s_\beta c_\beta) }{ t_\beta s_\beta^2 }\,,\nonumber\\
v^2 \lambda_3 &=&  m_h^2 + 2 m_{H^{\pm}}^2 - 2m_H^2 -  \frac{t_\beta (m_{12}^2 -m_H^2  s_\beta c_\beta)}{  s_\beta^2}\,,\nonumber\\
v^2 \lambda_4 &=&  m_A^2-  2 m_{H^{\pm}}^2 + m_H^2+  \frac{t_\beta (m_{12}^2 -m_H^2  s_\beta c_\beta)}{  s_\beta^2}\,,\nonumber\\
v^2 \lambda_5 &=&  m_H^2 - m_A^2+  \frac{t_\beta (m_{12}^2 -m_H^2  s_\beta c_\beta)}{  s_\beta^2} \,.
\label{eqn:poten-cba0}
\end{eqnarray}

In the type-I 2HDM, all the SM quarks and leptons couple exclusively to one Higgs doublet, which alters the Yukawa interactions of the neutral Higgs states relative to the SM,
\bea
\label{hffcoupling}
&&y_{h}^{f}=\left[\sin(\beta-\alpha)+\cos(\beta-\alpha)\kappa_f\right], \nonumber\\
&&y_{H}^{f}=\left[\cos(\beta-\alpha)-\sin(\beta-\alpha)\kappa_f\right], \nonumber\\
&&y_{A}^{f}=-i\kappa_f~(\text{for~}u),~~~~y_{A}^{f}=i \kappa_f~(\text{for~}d,~\ell),
\eea
where $\kappa_d=\kappa_\ell=\kappa_u \equiv 1/\tan\beta$ for the type-I 2HDM, and the generation indices are suppressed.
The interaction terms between the charged Higgs and the SM fermions are
\begin{align} \label{eqn:Yukawa2}
 \mathcal{L}_Y & = - \frac{\sqrt{2}}{v}\, H^+\, \Big\{\bar{u}_i \left[\kappa_d\,(V_{\text{CKM}})_{ij}~ m_{dj} P_R
 - \kappa_u\,m_{ui}~ (V_{\text{CKM}})_{ij} ~P_L\right] d_j + \kappa_\ell\,\bar{\nu} m_\ell P_R \ell
 \Big\}+\text{h.c.},
 \end{align}
where $i,j=1,2,3$.
The couplings of the neutral Higgs bosons to the SM gauge bosons, relative to their SM values, are given by,
\beq
y^{V}_h=\sin(\beta-\alpha),~~~
y^{V}_H=\cos(\beta-\alpha),\label{eqn:hvvcoupling}
\eeq
where $V=Z,~W$.
On the other hand, the couplings between the pseudoscalar $A$ and the SM gauge bosons are forbidden by CP conservation.

The interaction terms between one gauge boson and two scalar fields are
\begin{align}
\mathcal{L}_{SSV} = &\frac{g}{2}W^+_\mu\left((H^-\overset{\leftrightarrow}{\partial}^\mu A)-i\cos(\beta-\alpha)(H^-\overset{\leftrightarrow}{\partial}^\mu h)+i\sin(\beta-\alpha)(H^-\overset{\leftrightarrow}{\partial}^\mu H)+h.c. \right)\notag\\
&+\frac{g}{2c_W}Z_\mu\left(\cos(\beta-\alpha)(A\overset{\leftrightarrow}{\partial}^\mu h)-\sin(\beta-\alpha)(A\overset{\leftrightarrow}{\partial}^\mu H)\right)\notag\\
&+\left(ie\gamma_\mu+i\frac{g(c_W^2-s_W^2)}{2c_W}Z_\mu \right)(H^\mp\overset{\leftrightarrow}{\partial}^\mu H^\pm).
\end{align}
Here, we employ the abbreviations $c_W \equiv \cos\theta_W$ and $s_W \equiv \sin\theta_W$, with $\theta_W$ being the Weinberg angle.

\section{Relevant theoretical and experimental constraints}
\label{sec:currentlimits}

In this work, the light CP-even state $h$ is identified with the observed Higgs boson with a mass of 125 GeV.
We restrict ourselves to the near-alignment limit, $\cos(\beta-\alpha)\to 0$, under which the tree-level interactions of 
$h$ with the SM particles closely reproduce their SM predictions.
If $\tan\beta$ is sufficiently large or $A$ is light enough, the pseudoscalar $A$ is an LLP.
In the parameter region we consider, characterized by large values of $\tan\beta$ and tiny values of $\cos(\beta - \alpha)$, the Yukawa couplings of $h$ are approximately rescaled by a factor of $\sin(\beta - \alpha)$ (see Eq.~\eqref{hffcoupling}), and its couplings to the SM gauge bosons are likewise rescaled by $\sin(\beta - \alpha)$ (see Eq.~\eqref{eqn:hvvcoupling}).
Since $\sin(\beta - \alpha)$ is very close to unity in this region, these deviations are negligible.
Consequently, the corrections to the tree-level decay channels $h \to ZZ^*, WW^*$, and $h \to f\bar{f}$ can be safely neglected.
The charged Higgs bosons can contribute to the $h\to \gamma\gamma$ process through loop effects~\cite{Wang:2013sha}.
In addition, the decay mode $h\to AA$ enhances the total width of $h$, and thus affects Br$(h\to \gamma\gamma)$. 
As a result, measurements of the diphoton signal strength can impose stringent limits on the relevant parameter space.
Among the measured channels, the diphoton mode currently provides the most precise determination of the properties of the 125 GeV Higgs boson through its signal strength~\cite{ParticleDataGroup:2020ssz}
\beq
\mu_{\gamma\gamma}= 1.11^{+0.10}_{-0.09}.
\eeq

Alternatively, we employ $\texttt{HiggsTools}$~\cite{Bahl:2022igd} to compute the total $\chi^2$ based on the latest LHC measurements of the signal strength of the 125 GeV Higgs boson.
The required \texttt{SLHA}~\cite{Allanach:2008qq} file is generated with $\texttt{SPheno v4.0.5}$~\cite{Porod:2003um} and subsequently provided as input to $\texttt{HiggsTools}$.
We pay particular attention to the parameter points with $\chi^2-\chi^2_{\rm min} \leq 6.18$, where $\chi^2_{\rm min}$ denotes the minimum of $\chi^2$.
These points lie within the $2\sigma$ range in any two-dimensional plane of the model parameters, for explaining the Higgs data.

For $m_A < 62.5$ GeV, the SM-like Higgs boson can decay into a pair of long-lived $A$'s, which subsequently decay into hadronic jets, producing a pair of displaced vertices. This signature has been actively investigated at the LHC.
Based on the ATLAS analyses in Refs.~\cite{ATLAS:2022gbw,ATLAS:2025pak}, we adopt the bound Br$(h\to AA) < 0.1\%$, which corresponds to the most stringent limit reported from these ATLAS searches for a pair of displaced vertices arising from LLPs decaying into hadronic jets.
We note that for relatively small values of $\tan\beta$ and large values of $m_A$, the $A$ particle can be less long-lived, and in this regime, prompt limits on the branching ratio of the Higgs-boson decays into multiple jets via BSM light states may apply.
However, such bounds, currently with the existing LHC Run~2 data, are quite loose and only at the level of~$\mathcal{O}(10\%)$~\cite{ATLAS:2021ypo,Cepeda:2021rql,Castillo:2022bin,CMS:2024zfv}.
Therefore, we do not take them into account and work with the stringent bound of Br$(h\to A A)<0.1\%$ mentioned above.

Our analysis also incorporates theoretical requirements including vacuum stability, perturbative behavior, and unitarity. 
Firstly, to ensure perturbativity, each quartic coupling in the Higgs potential is required to remain below $4\pi$.
Second, the requirements ensuring vacuum stability are given by~\cite{Deshpande:1977rw}
\beq
\label{eqn:vacuum-condi}
\lambda_1 > 0,~~\lambda_2 > 0,~~\lambda_3 + \sqrt{\lambda_1\lambda_2} > 0,~~\lambda_3 + \lambda_4 - \mid\lambda_5\mid +\sqrt{\lambda_1\lambda_2} > 0.
\eeq

Furthermore, perturbative unitarity at tree level requires that the eigenvalues of the scattering matrix for $2 \to 2$ 
processes involving scalar quartic couplings should not exceed $8\pi$.
 This criterion places constraints on the parameters $\lambda_{1,2,3,4,5}$ \cite{Kanemura:1993hm,Akeroyd:2000wc},
\begin{eqnarray} \label{eqn:unitarity}
|a_{\pm}|, |b_\pm|, |c_\pm|, |{ e}_\pm|, |{ f}_\pm|, |{ g}_\pm|
\,\le\, 8\pi \,,
\end{eqnarray}
with 
\begin{eqnarray}
a_\pm^{} &=& \tfrac{3}{2}(\lambda_1+\lambda_2) \pm \sqrt{\tfrac{9}{4}(\lambda_1-\lambda_2)\raisebox{0.3pt}{$^2$}+(2\lambda_3+\lambda_4)^2} \,, \nonumber\\
b_\pm^{} &=& \tfrac{1}{2}(\lambda_1+\lambda_2) \pm
\sqrt{\tfrac{1}{4}(\lambda_1-\lambda_2)\raisebox{0.3pt}{$^2$}+\lambda_4^2} \,,\nonumber \\
c_\pm^{} \,&=&\, \tfrac{1}{2}(\lambda_1+\lambda_2) \pm
\sqrt{\tfrac{1}{4}(\lambda_1-\lambda_2)\raisebox{0.3pt}{$^2$}+\lambda_5^2} \,, \nonumber\\
{ e}_\pm^{} &=& \lambda_3^{} + 2 \lambda_4^{} \pm 3 \lambda_5^{} \,, \nonumber\\
{ f}_\pm^{} \,&=&\, \lambda_3^{} \pm \lambda_4^{} \,,\nonumber\\
{ g}_\pm \,&=&\, \lambda_3^{} \pm \lambda_5^{} \,.
\end{eqnarray}

Additional loop contributions to the electroweak oblique parameters (S, T, and U) can arise from the exchange of 
the extra Higgs states in gauge-boson self-energy diagrams.
We evaluate these quantities with the \texttt{2HDMC} package~\cite{Eriksson:2009ws}, and compare them with the global-fit values extracted from Ref.~\cite{ParticleDataGroup:2024cfk},
\beq
S=-0.04\pm 0.10,~~  T=0.01\pm 0.12,~~ U=-0.01 \pm 0.09,
\eeq
accompanied by the correlation coefficients
\beq
\rho_{ST} = 0.93,~~  \rho_{SU} = -0.70,~~  \rho_{TU} = -0.87.
\eeq

For the numerical analysis, we perform a random scan over $\cos(\beta-\alpha)$, $\tan\beta$, $m_H$, $m_{H^\pm}$, and $m_A$, within the following parameter ranges,
\bea
\label{eqn:space}
&&0 \leq \cos(\beta-\alpha) \leq 10^{-2},~~5\times 10^3 \leq \tan\beta \leq 10^6,\nonumber\\
&&200~ {\rm GeV} \leq m_H=m_{H^\pm} \leq 800~ {\rm GeV},~~10~ {\rm GeV} \leq m_A \leq 100\text{ GeV}.
\eea
If the CP-even mixing angle is shifted as $\alpha \to \alpha + \pi$, $\cos(\beta-\alpha)$ will change its sign and become negative.
This transformation corresponds to a simultaneous field redefinition $h\to -h$ and $H \to -H$,  which leaves all physical observables unaffected.
The condition $m_H=m_{H^\pm}$ is preferred, considering constraints from the oblique parameters (see also Refs.~\cite{Gerard:2007kn,deVisscher:2009zb}).
To ensure that the decay channels $H\to AZ$ and $H^{\pm}\to AW^{\pm}$ remain open even for a relatively light $H$, we choose $m_A\leq 100$ GeV.
Furthermore, for a heavier $A$, the decay modes $A\to h^{(*)}Z$ are enhanced and thus the total decay width of $A$ increases.
Since the partial decay width of $A\to h^{(*)}Z$ is not suppressed for large values of $\tan\beta$, this complicates the discussion on the lifetime of $A$.

From the first relation in eq.~\eqref{eqn:poten-cba0}, the perturbativity bound on $\lambda_1$ tends to favor $m^2_{12}- m_H^2 s_\beta c_\beta \to 0$ for huge values of $\tan\beta$.
Nonetheless, if $\cos(\beta-\alpha)=0$ and $m^2_{12}- m^2_H s_\beta c_\beta =0$, simultaneously solving eq.~\eqref{eqn:poten-cba0} and the final vacuum-stability condition given in eq.~\eqref{eqn:vacuum-condi} imposes a stringent constraint: $m_h^2+m_A^2> m_H^2$, which conflicts with a light $m_A$.
Therefore, to satisfy the theoretical constraints, the relevant parameters must be carefully tuned, with fine-tuning becoming more severe at higher $\tan\beta$.
Correspondingly, the quantity $m^2_{12}-m^2_H s_\beta c_\beta$ will serve later in a definition of the degree of fine-tuning; see Eq.~\eqref{eqn:Rf} below.

\begin{figure}[t]
\centering
\includegraphics[width=0.9\textwidth]{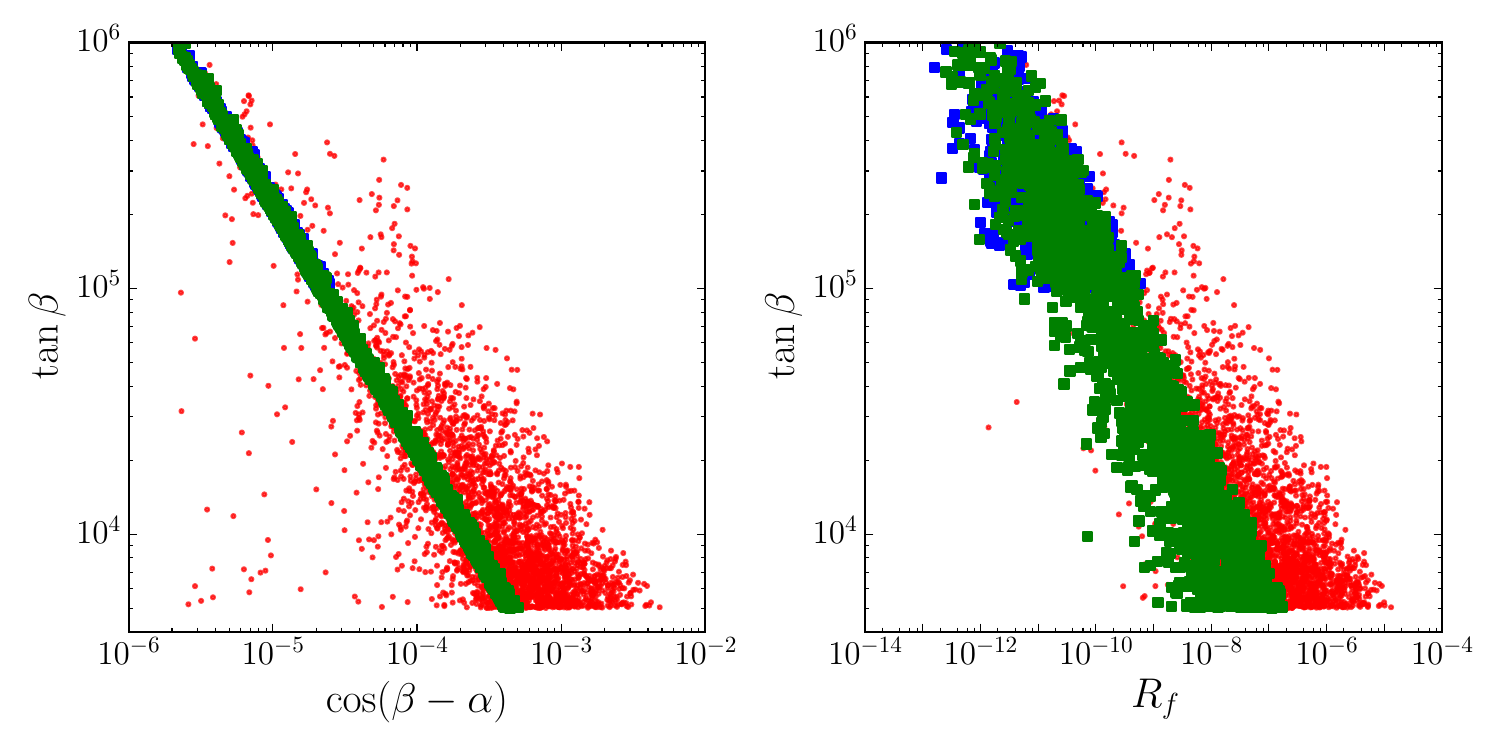}
\caption{All the points that satisfy the theoretical requirements and the oblique parameter bounds, for $10\text{ GeV}<m_A< 62.5\text{ GeV}$, shown in the $\tan\beta$ vs.~$\cos{(\beta-\alpha)}$ plane (left) and the $\tan\beta$ vs.~$R_f$ plane (right). The green points are allowed by the joint constraints of Br$(h\to AA) < 0.1\%$ and the diphoton signal data of the 125 GeV Higgs, while the blue and red points are excluded. In contrast, both the green and blue points are accommodated by the requirement $\chi^2-\chi^2_{\text{min}} \leq 6.18$ and the red points are forbidden.}
\label{figxthe}
\end{figure}

\begin{figure}[t]
\centering
\includegraphics[width=0.9\textwidth]{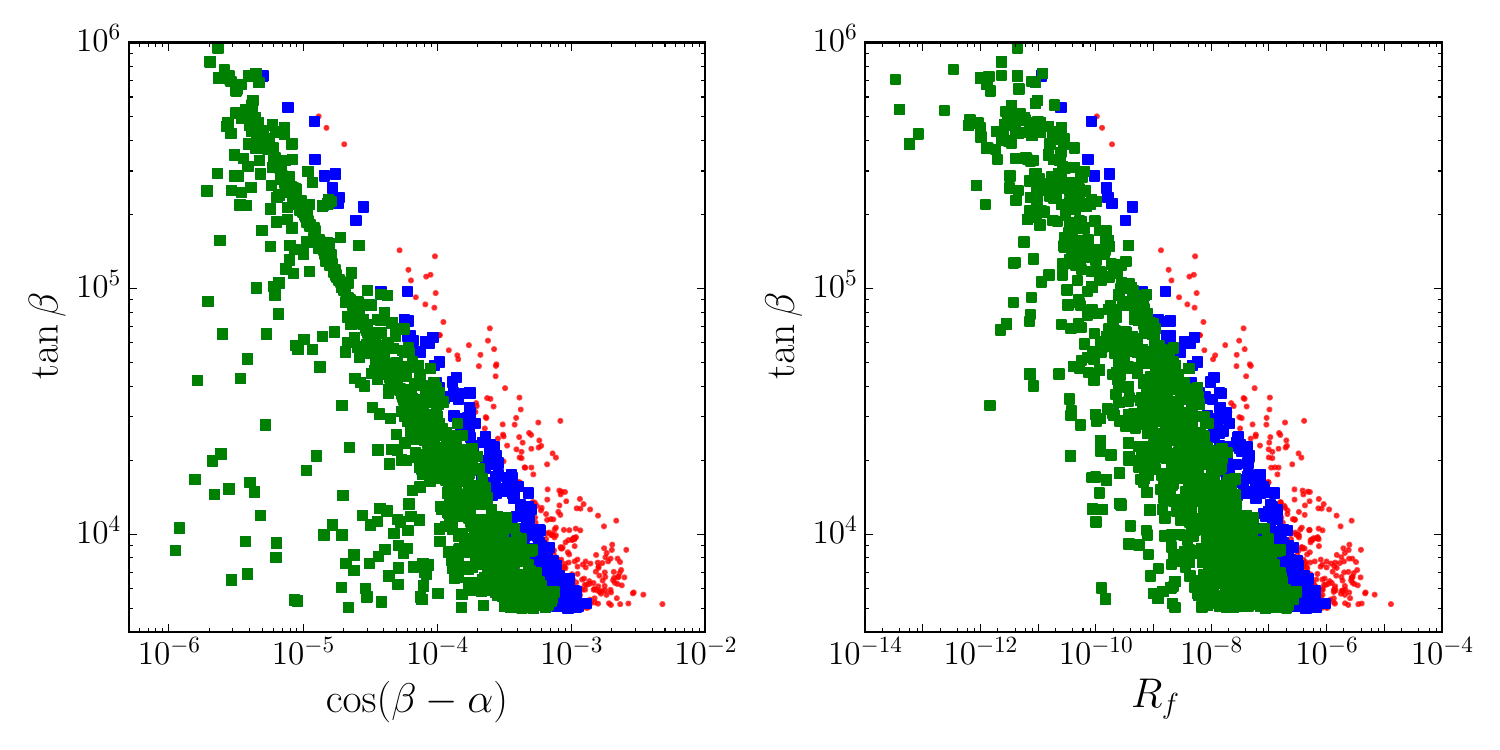}
\caption{The same as figure~\ref{figxthe}, but for $62.5\text{ GeV}<m_A<100\text{ GeV}$.}
\label{figbthe}
\end{figure}

\begin{figure}[t]
\centering
\includegraphics[width=0.9\textwidth]{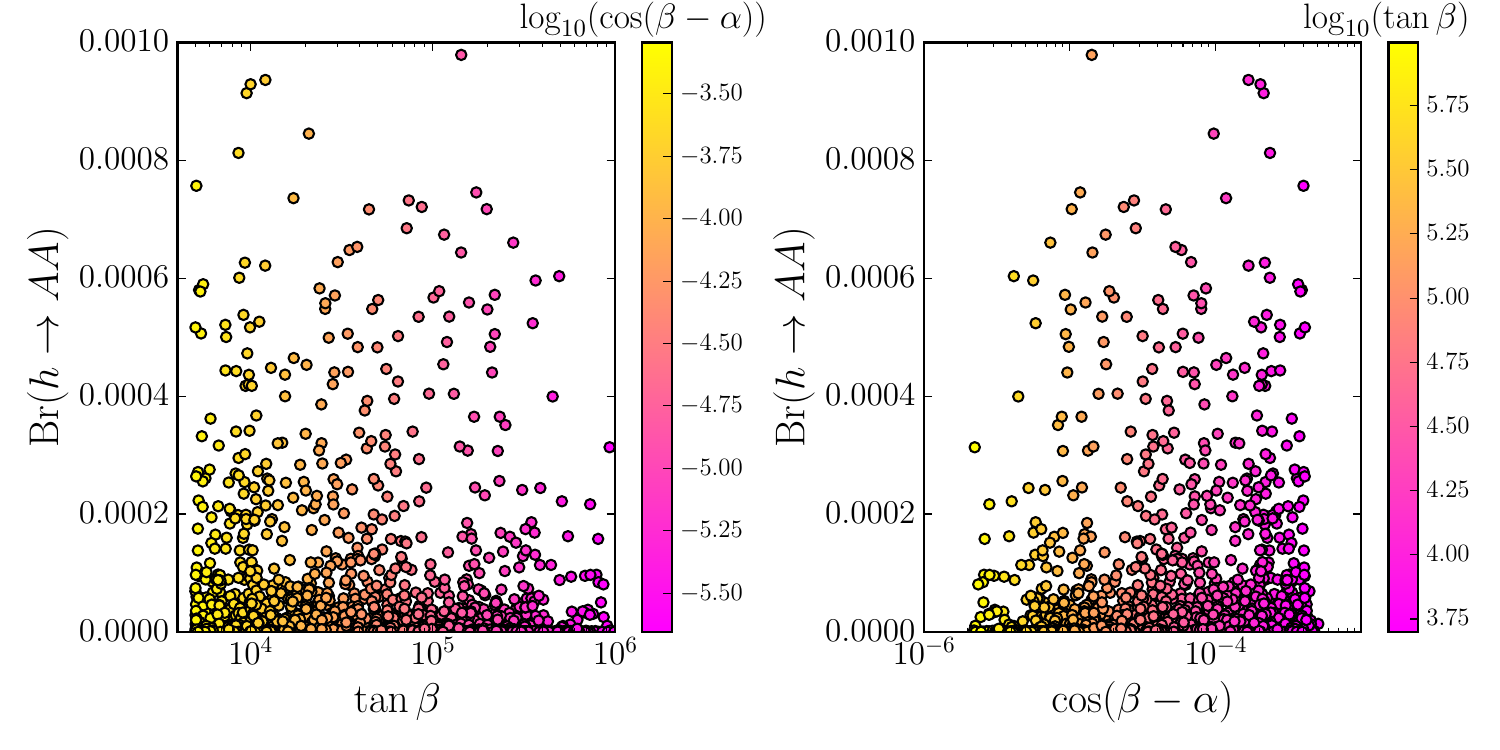}
\caption{Br($h\to AA$) as functions of $\tan\beta$ and $\cos(\beta-\alpha)$, for 10 GeV $<m_A<$ 62.5 GeV.
All the displayed points satisfy the theoretical requirements, the oblique parameter bounds, and the diphoton signal data of the 125 GeV Higgs.}
\label{figxbr}
\end{figure}

After applying the constraints from theory, the oblique-parameter, Br$(h\to AA) < 0.1\%$, the diphoton signal data of the 125 GeV Higgs, the surviving parameter points are displayed in figure~\ref{figxthe} for $10\text{ GeV}<m_A< 62.5\text{ GeV}$ and in figure~\ref{figbthe} for $62.5\text{ GeV}<m_A<100\text{ GeV}$.
In both figures, two plots are displayed, in the $\tan\beta$ vs.~$\cos(\beta-\alpha)$ and the $\tan\beta$ vs.~$R_f$ planes, respectively.
The parameter $R_f$ is defined to be 
\begin{eqnarray}
R_f\equiv\frac{\mid m^2_{12}- m^2_H s_\beta c_\beta\mid}{m^2_{12}},   \label{eqn:Rf}
\end{eqnarray}
which quantifies the degree of fine-tuning required between $m_{12}^2$ and $m^2_H s_\beta c_\beta$.
Here, $m_{12}^2$ is used in the denominator since it is an independent input parameter of the Higgs potential, while $m_H^2 s_\beta c_\beta$ is a derived combination of physical quantities; in the tuned regime where $m_{12}^2 \simeq m_H^2 s_\beta c_\beta$, the numerical value of $R_f$ is insensitive to this normalization choice.

As shown in figure~\ref{figxthe} and figure~\ref{figbthe}, the green points are allowed by the joint constraints of Br$(h\to AA) < 0.1\%$ and the diphoton signal data of the 125 GeV Higgs~\cite{ParticleDataGroup:2020ssz}, while the blue and red points are excluded.
In contrast, the requirement $\chi^2-\chi^2_{\text{min}} \leq 6.18$ accommodates both the green and blue points and excludes the red points.
As illustrated in figure~\ref{figxthe}, for $10~\text{GeV}<m_A<62.5$~GeV, the allowed points obtained from the two approaches exhibit no visible differences. For $62.5~\text{GeV}< m_A < 100$~GeV, imposing the requirement $\chi^2-\chi^2_{\text{min}} < 6.18$ leads to slightly weaker constraints on $\cos(\beta-\alpha)$ and $\tan\beta$ compared to the alternative approach, as shown in figure~\ref{figbthe}.

As seen from the analysis above, for large values of $\tan\beta$, the theoretical constraints disfavor the possibility that $\cos(\beta-\alpha)=0$ and $m_{12}^2- m_H^2 s_\beta c_\beta=0$ hold exactly.
With increasing $\tan\beta$, $\cos(\beta-\alpha)$ should decrease so that the theoretical constraints are satisfied.
The left panels of figure~\ref{figxthe} and figure~\ref{figbthe} show that at $\tan\beta=10^6$, $\cos{(\beta-\alpha)}$ should be as small as in the order of $10^{-6}$.
The right panels of figure~\ref{figxthe} and figure~\ref{figbthe} indicate that $R_f$ tends to decrease for larger values of $\tan\beta$, reaching levels below $\order(10^{-12})$ for $\tan\beta=10^6$. 
This implies that the degree of fine-tuning increases significantly at larger values of $\tan\beta$, which in turn places more stringent demands on the numerical precision of computer-based calculations owing to the enhanced sensitivity to floating-point rounding effects.\footnote{For fine-tuning so severe that it exceeds machine precision, the computer-evaluated values of $m_{12}^2$ and $m_H^2s_\beta c_\beta$ develop unstable and unreliable trailing digits (least significant digits). As a result, the extracted values of $m_{12}^2$, $m_H^2s_\beta c_\beta$, and consequently $R_f$ are affected by numerical noise.}
We therefore restrict our discussion to $\tan\beta\lesssim 10^6$.

For $10\text{ GeV}<m_A< 62.5\text{ GeV}$, the decay channel $h\to AA$ is kinematically accessible, and the parameter space is hence subject to the bound Br$(h\to AA)<0.1\%$.
As a result, a significant portion of the parameter space allowed by theoretical requirements is excluded, leading to a strong correlation between $\cos{(\beta-\alpha)}$ and $\tan\beta$ in the allowed parameter space, as shown in the left panel of figure~\ref{figxthe}.
The surviving points are projected onto the Br$(h\to AA)$ vs.~$\tan\beta$ and the Br$(h\to AA)$ vs.~$\cos(\beta-\alpha)$ planes in figure~\ref{figxbr}.
Most of the parameter points tend to lie within the region where the values of Br$(h\to AA)$ are below $\sim0.04\%$.

For $62.5\text{ GeV}<m_A< 100\text{ GeV}$, the decay $h\to AA$ is kinematically forbidden.
Although one of the two $A$ bosons could be produced off-shell in Higgs boson decays, the corresponding width is safely negligible since its fermion couplings are strongly suppressed at very large $\tan\beta$.
As a result, the allowed parameter space in this mass range is broader than that in the case of $10\text{ GeV}<m_A< 62.5\text{ GeV}$, as illustrated in the left panel of figure~\ref{figbthe}.

\begin{figure}[t]
\centering
\includegraphics[width=0.5\textwidth]{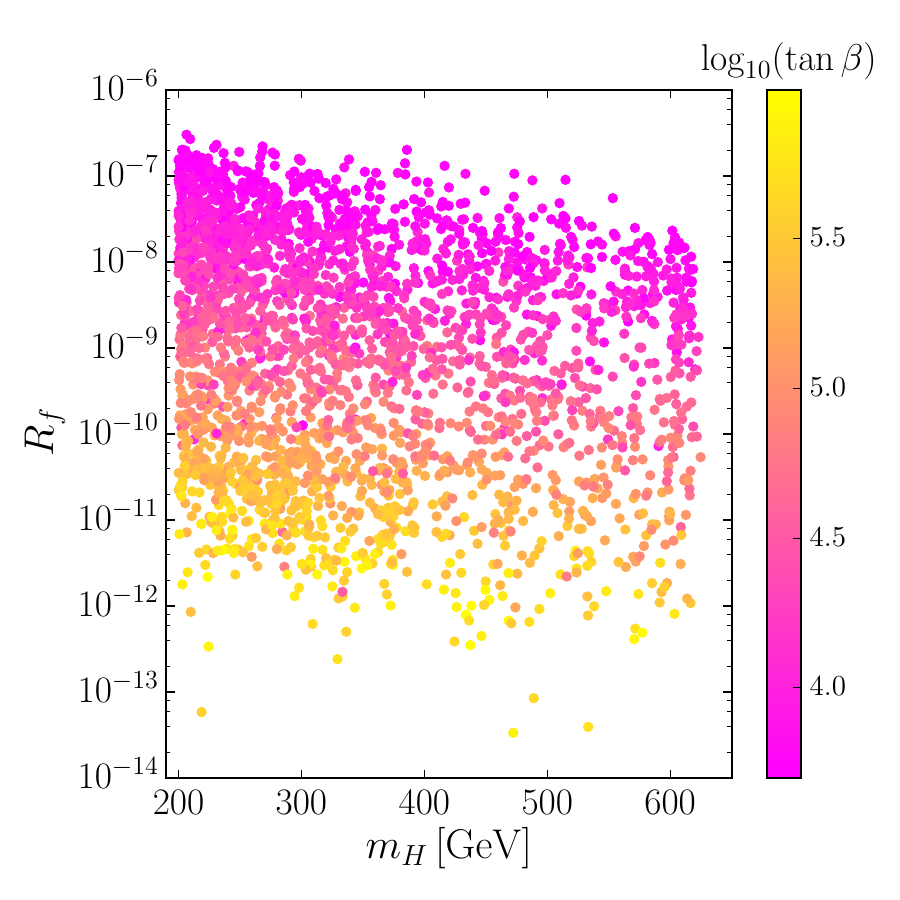}
\caption{All the points that satisfy the theoretical requirements, the oblique parameter bounds, Br$(h\to AA) < 0.1\%$, and the diphoton signal data of the 125 GeV Higgs, for $10\text{ GeV}<m_A< 100\text{ GeV}$.}
\label{figthemh2}
\end{figure}

We present an additional plot in figure~\ref{figthemh2}, displaying parameter points in the $(m_H, R_f)$ plane that satisfy all the considered theoretical and experimental constraints for $m_A$ between 10 GeV and 100 GeV.
For these points, we assign different colors for varying levels of $\tan\beta$.
The plot shows that $R_{f}$ tends to decrease with a heavier $H$.
Although we scan over the range of $200\text{ GeV}\leq m_H=m_{H^\pm} \leq 800\text{ GeV}$, the allowed parameter space only survives for $m_H=m_{H^\pm}\leq 630\text{ GeV}$. 
From eq.~\eqref{eqn:poten-cba0}, one observes that a large mass-splitting between $m_A$ and $m_H$ ($m_{H^\pm}$) would increase the values of $\lambda_4$ and $\lambda_5$, thereby challenging perturbativity and unitarity constraints.

\begin{figure}[t]
\centering
 \includegraphics[height=0.45\textwidth]{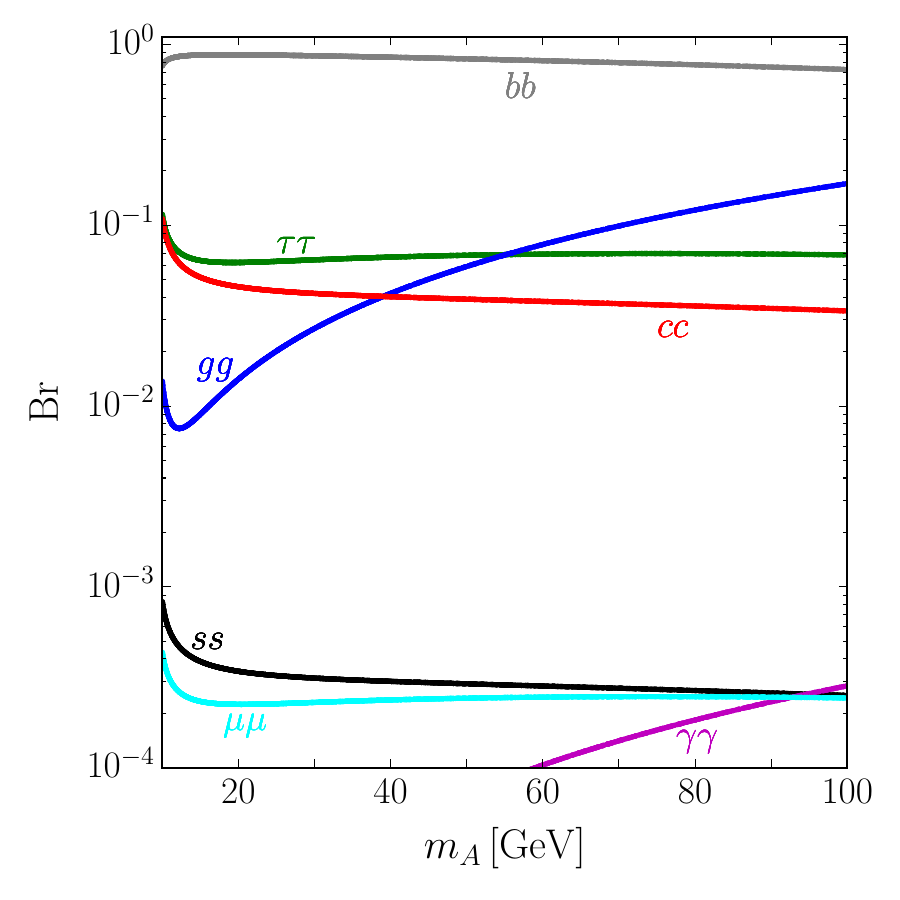}
 \includegraphics[height=0.50\textwidth]{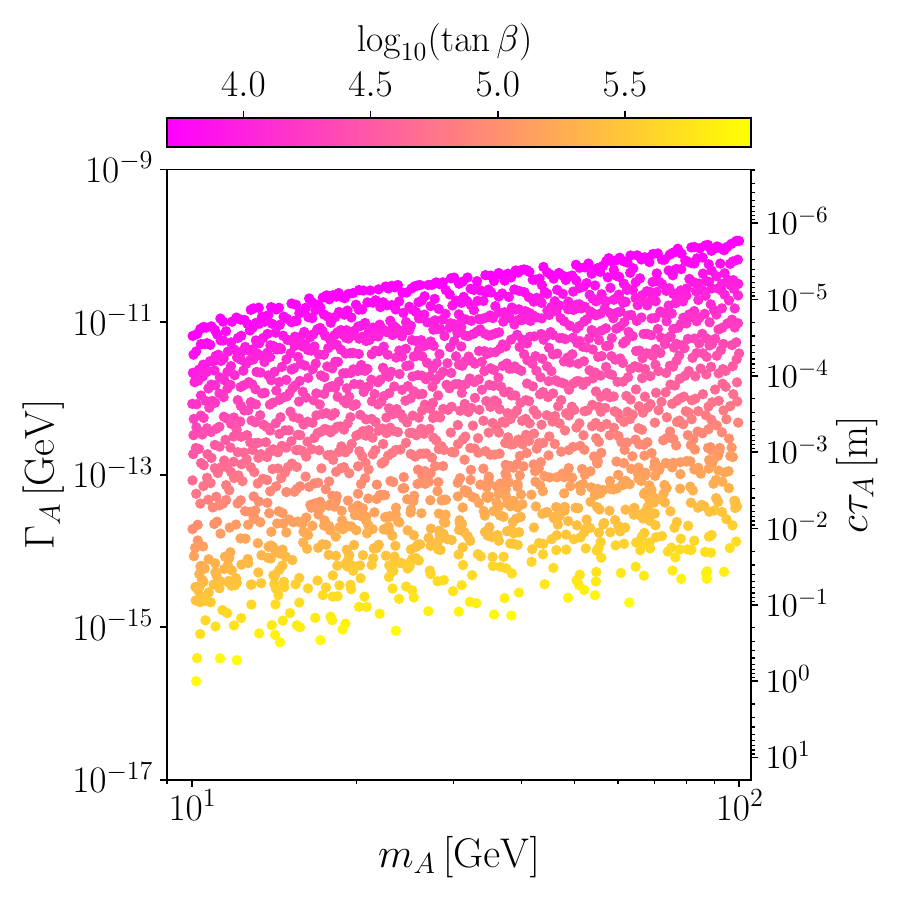}
\caption{Left panel: Branching ratios of various decay channels of $A$.
Right panel: The total decay width of $A$ vs.~$m_A$ satisfying the constraints of theory and the oblique parameters, Br$(h\to AA) < 0.1\%$, and the diphoton signal data of the 125 GeV Higgs. The corresponding values of $c\tau_A$ are labeled on the right $y$-axis.}
\label{fig:br_wid}
\end{figure}

From eq.~\eqref{eqn:hvvcoupling}, one sees that the couplings of $H$ to the SM gauge bosons are proportional to $\cos(\beta-\alpha)$, which are strongly suppressed in the parameter region allowed by both Br$(h\to AA)< 0.1\%$ and the diphoton signal data of the 125 GeV Higgs,  where $\cos(\beta-\alpha)<10^{-3}$.
The Yukawa couplings of the additional Higgses ($H$, $H^\pm$, $A$) are highly suppressed owing to the tiny values of $1/\tan\beta$.
Consequently, the production cross sections of these extra Higgs bosons at the LHC, through gluon-gluon fusion, vector-boson fusion, or associated production with fermions or electroweak gauge bosons, are all significantly reduced.
We employ $\texttt{HiggsBounds}$~\cite{Bechtle:2020pkv,Bechtle:2008jh} through \texttt{HiggsTools}~\cite{Bahl:2022igd} to implement the exclusion limits from collider searches for neutral and charged Higgs states at 95\% confidence level (C.L.), and our results show that all parameter points compatible with the 125 GeV Higgs signal data remain allowed.

We comment now on the potential contributions of the considered type-I 2HDM scenarios to the total decay width of the $Z$- and $W$-bosons.
We note that in the type-II 2HDM, the bottom-quark Yukawa couplings of the additional Higgs bosons $(H, H^{\pm}, A)$ are strongly enhanced at large $\tan\beta$, and the resulting corrections to $Z\to b\bar{b}$ should therefore be taken into account.
In the lepton-specific 2HDM, the tau-lepton Yukawa couplings are similarly enhanced, requiring the inclusion of the model’s corrections to $Z\to \tau\bar{\tau}$.
In contrast, in the type-I 2HDM, which is the focus of our paper, all fermionic Yukawa couplings of the additional Higgs states are highly suppressed with large values of $\tan\beta$; as a result, the model’s contributions to the total decay widths $Z$ and $W^{\pm}$ can be safely neglected.

The interactions of the pseudoscalar $A$ with the SM leptons and quarks scale inversely with $\tan\beta$ at the amplitude level.
Consequently, if $\tan\beta$ is sufficiently large, $A$ can acquire a long lifetime. 
The primary decay channels of $A$ depend on its mass.
Using the \texttt{2HDMC} package, we compute the decay widths of $A$ for masses between 10 GeV and 100 GeV.
The corresponding branching ratios for various channels are presented in the left panel of figure~\ref{fig:br_wid}.
In this mass window, the $A$ particle predominantly decays into $b\bar{b}$, with the corresponding branching ratio ranging between $70\%$ and $90\%$ approximately.

The total width of $A$ is displayed in the right panel of figure~\ref{fig:br_wid}, shown in the $\Gamma_A$ vs.~$m_A$ plane for various levels of $\tan\beta$.
Incidentally, $c\tau_A$ is correspondingly plotted.
$\Gamma_A$ increases for a heavier $A$ or a smaller $\tan\beta$.
Notably, for sufficiently small $\Gamma_A$, the corresponding decay length, $c\tau_A$, can exceed the meter scale.
We should also mention that $\Gamma_A$ is essentially independent of $m_H$.

Before ending the section, we remark that a pseudoscalar with a mass above 10~GeV is not subject to existing experimental constraints on a light pseudoscalar, which include supernova~\cite{Turner:1987by,Ellis:1987pk}, searches for axion-like particles from CHARM~\cite{CHARM:1985anb,Gorbunov:2021ccu}, $B$-meson decays~\cite{LHCb:2015nkv,LHCb:2016awg} and $D$-meson decays~\cite{LHCb:2020car} from LHCb, kaon decays from NA62~\cite{NA62:2021zjw}, MicroBooNE~\cite{MicroBooNE:2021sov}, and E949~\cite{BNL-E949:2009dza}.

\section{Experimental analysis}
\label{sec:analysis}

\begin{figure}[tb]
\centering
\includegraphics[width=0.9\textwidth]{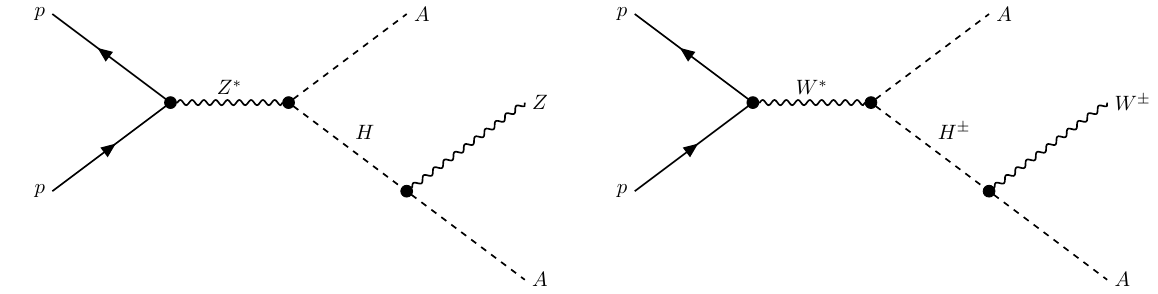}
\caption{The Feynman diagrams for the dominant production modes of $A$ via electroweak processes at the LHC.} \label{figfeynman}
\end{figure}

Since the couplings of $A$ with the SM fermions are strongly suppressed by the factor $1/\tan\beta$, the production of $A$ at the LHC proceeds predominantly in electroweak processes including one charged-current and one neutral-current process,
\begin{align}
pp\to & W^{\pm *} \to H^\pm A \to W^\pm AA, \label{process1}\\
pp\to &       Z^* \to HA\to ZAA, \label{process2}
\end{align}
and the corresponding Feynman diagrams are shown in figure~\ref{figfeynman}.
Here, we note that $H^\pm\to W^\pm\,A$ and $H\to Z\,A$ are the primary decay channels of $H^\pm$ and $H$, respectively, for $m_H=m_{H^\pm}>>m_A$. 

Since all of $W$-boson, $Z$-boson, and the pseudoscalar particle $A$ in the mass range of our interest, decay primarily into hadronic final states, we choose to focus on the final states of $W, Z\to jj$ and $A\to b\bar{b}$, where $j$ denotes a quark-jet including the following quarks: $u, d, c, s, \text{ and }b$, as well as their corresponding anti-particle counterparts.
Since the $A$ particle is long-lived in the large region of parameter space (see figure~\ref{fig:br_wid}), we study the characteristic signature of displaced vertices (DVs) accompanied with multiple jets at the LHC.

This signature has been searched for recently at the ATLAS experiment with the full Run-2 data of $139$~fb$^{-1}$~\cite{ATLAS:2023oti} which targets at long-lived electroweakinos in the R-parity-violating supersymmetry.
While no discovery of BSM physics was made, stringent bounds on the considered model were obtained.
The search has been recast in Ref.~\cite{Cheung:2024qve} according to ATLAS instructions~\cite{atlas_recast_instruction},\footnote{The recast code can be accessed at the public LLP Recasting Repository~\url{https://github.com/llprecasting/recastingCodes/}. A second version of the recast~\cite{Heisig:2024xbh} is also provided in the same repository.} and the recast has been applied in Ref.~\cite{Wang:2024ieo} for deriving corresponding bounds on several benchmark models of LLPs pair-produced in exotic Higgs boson decays.
Furthermore, Ref.~\cite{Beltran:2025ilg} studied a theoretical scenario with long-lived heavy neutral leptons and effective operators involving the SM top quarks, where the recast search was found to fail constraining the scenario, mainly as a result of its too strong jet-$p_T$ thresholds.
The authors of Ref.~\cite{Beltran:2025ilg} thus proposed a search for the same signature, combining the relatively low jet-$p_T$ thresholds taken in a past 8-TeV search for DVs at ATLAS~\cite{ATLAS:2015oan} and the DV-reconstruction approach applied in Ref.~\cite{ATLAS:2023oti}.\footnote{We note that Ref.~\cite{Beltran:2025ilg} pointed out that the lowering of the jet-$p_T$ thresholds should, in principle, not raise the background level, because background suppression is mainly achieved via DV reconstruction.\label{footnote:jet_pt_background}}
We will test our theoretical scenario against both of these two analyses, referred to as ``original analysis'' and ``modified analysis'', respectively.

For both analyses, we briefly describe below the event-selection procedures.
The two searches are implemented in \texttt{C++} code including both acceptance requirements and parameterized efficiencies.
The latter, provided by the ATLAS collaboration~\cite{atlas_recast_instruction}, are for accounting for delicate requirements that are difficult to simulate with Monte Carlo (MC) tools, such as multi-jet trigger and material effects.
The acceptance requirements are imposed at two sequential levels: event and vertex.
The event-level acceptance concerns solely selections on truth-jet's transverse momentum while the vertex-level acceptance pertains to selection of displaced vertices of \textit{high quality}, to be explained below.

Here, for reconstructing truth-level jets we have implemented a toy-detector module in \texttt{Pythia8}~\cite{Sjostrand:2014zea,Bierlich:2022pfr}, following the ATLAS recast instructions of this search~\cite{atlas_recast_instruction}.
Concretely, we reconstruct truth-level jets with the \texttt{FastJet}~\cite{Cacciari:2011ma} package, exploiting the anti-$k_t$ algorithm with $R = 0.4$ where neutrinos and muons are excluded.
We emphasize that here the jet definition includes both prompt and displaced jets.
We have also modeled detector response for the measurement of jet $p_T$, considering detector acceptance, resolution, and smearing on their transverse momenta, following Ref.~\cite{Allanach:2016pam}.

\begin{table}[t]
\begin{center}
\begin{tabular}{c|c|c}
Analysis            & original                                                  & modified                                                              \\ \hline
              & $n^{137}_{\text{jet}}\geq 4$ or $n^{101}_{\text{jet}}\geq 5$    & $n^{90}_{\text{jet}}\geq 4$ or $n^{65}_{\text{jet}}\geq 5$                   \\
Event-level acceptance & or $n^{83}_{\text{jet}}\geq 6$ or $n^{55}_{\text{jet}}\geq 7$, & or $n^{55}_{\text{jet}}\geq 6$                 \\
       &      $n^{70}_{\text{displaced jet}}\geq 1$ or $n^{50}_{\text{displaced jet}}\geq 2$                                                           &  \\ \hline
\end{tabular}
\caption{Event-level acceptance, imposing selection cuts on the transverse momentum of the truth-jets. Here, $n^{137}_{\text{jet}}$ refers to the number of jets with a $p_T$ at least 137 GeV, and similarly for the other notations.}
\label{tab:jet_pt_requirement}
\end{center}
\end{table}

\subsection{The ``original'' analysis}

The ATLAS search of the original analysis~\cite{ATLAS:2023oti}, with an integrated luminosity of $139\text{~fb}^{-1}$ employs two signal regions (SRs), called ``High-$p_T$ jet'' and ``Trackless jet'',\footnote{In this experimental search, a jet is deemed trackless ``if the sum of the $p_T$ of all standard (non-LRT) tracks associated with the jet is less than 5 GeV''~\cite{ATLAS:2023oti}. This selection is not explicitly taken in the recast though, as the recast follows closely the instructions provided by the ATLAS collaboration instead.} respectively.
In our theoretical scenarios, we will focus on the latter SR since it shows stronger constraining power than the former for all the relevant parameter points in this study.

In the practice of the recast, we first apply the jet-$p_T$ selection cuts, requiring at least 4, 5, 6, or 7 jets to have a $p_T$ larger than or equal to 137, 101, 83, or 55 GeV, respectively.
Additionally, the SR requires certain numbers of displaced jets that should be sufficiently hard, where a ``displaced jet'' should be a jet that is determined to have originated from the decay of an LLP by checking $\Delta R$ between the LLP’s decay products and the truth jet.
These are summarized in table~\ref{tab:jet_pt_requirement}.

The events that have passed these jet-$p_T$ selections should subsequently be checked against a set of requirements on the displaced vertices.
Concretely speaking, at least one DV (corresponding to an LLP decay) in the events should fulfill the following requirements:
\begin{enumerate}
    \item Fiducial volume: $4\text{~mm}<R_{xy}<300\text{~mm}$ and $|z|<300\text{~mm}$, with $R_{xy}$ and $|z|$ being, respectively, the absolute distance of the vertex to the IP in the transverse and longitudinal directions.
    \item Transverse impact parameter: the DV should have at least one associated track with a sufficiently large absolute transverse impact parameter: $|d_0|>2\text{~mm}$.
    \item Selected decay products: the DV should have at least 5 massive decay products satisfying the following two conditions:
    \begin{enumerate}
        \item The decay product should be a track with a transverse decay length in the laboratory frame ($\beta_T \gamma c\tau$) larger than 520 mm, with $\beta_T$ denoting its absolute speed in the transverse direction, $\gamma$ being the Lorentz boost factor, and $c\tau$ being the decay length in its rest frame.
        \item The transverse momentum $p_T$ and the electric charge $q$ of the decay product should obey the relation $p_T/|q|> 1\text{~GeV}$.
    \end{enumerate}
    \item DV invariant mass:  $m_{\text{DV}}> 10$ GeV, where $m_{\text{DV}}$ is evaluated with the decay products passing the conditions above and these decay products are assumed to have the mass of a charged pion.
\end{enumerate} 

On the events that have passed both event- and vertex-level acceptance requirements, we apply the above-mentioned, parameterized efficiency functions provided by ATLAS.
The event-level parameterized efficiencies are functions of the truth-jet scalar $p_T$ sum and $R_{xy}$ of the furthest LLP decay, and the vertex-level parameterized efficiencies take input parameters of $R_{xy}$, $m_\text{DV}$, and LLP decay-product multiplicity.
In particular, the latter estimate the efficiencies of DV reconstruction.

Primary sources of background events include erroneous merge of nearby DVs of small invariant masses by vertexing algorithms giving rise to a large $m_{\text{DV}}$, particles' hadronic interactions with detector materials, and accidental crossings between a track and unrelated low-mass DVs.
However, after the selection cuts above have been applied, order-1 or even lower background levels are expected.
Specifically, the background-event number was estimated to be $0.83^{+0.51}_{-0.53}$ in this SR and 0 event was observed in the real data analysis.

In addition, to make sensitivity projections for the HL-LHC with an integrated luminosity of 3000 fb$^{-1}$, we assume future advancements in technologies and analysis algorithms can contain the background levels under control even though more severe pileup contamination is expected at the HL-LHC.
Therefore, we will assume vanishing background level for an integrated luminosity of $3000$~fb$^{-1}$, too.

\subsection{The ``modified'' analysis}

For the modified analysis, as explained above, the jet-$p_T$ thresholds are lowered compared to the original analysis, which is inspired by a past ATLAS 8-TeV search for DVs~\cite{ATLAS:2015oan}.
Concretely, the search now requires at least 4, 5, or 6 truth-jets that should have a transverse momentum of at least 90, 65, or 55~GeV, respectively (see table~\ref{tab:jet_pt_requirement}).
Improved event-level acceptance guarantees stronger cutflow efficiencies.

For the events having passed these jet-$p_T$ selections, the same set of vertex-level acceptance criteria as those in the trackless-jet SR in the original analysis from Ref.~\cite{ATLAS:2023oti} are imposed.

Finally, we apply (only) the vertex-level parameterized efficiencies provided by the ATLAS collaboration in the search for DVs and multiple jets~\cite{ATLAS:2023oti}.

As explained in footnote~\ref{footnote:jet_pt_background}, the suppression of background events is mainly achieved by the DV selection and reconstruction.
Therefore, with this search, we also assume 0 background event for an integrated luminosity of both 139~fb$^{-1}$ and 3000~fb$^{-1}$.

\subsection{Computation procedure}

Now we explain our numerical-computation procedures.
We first implement a UFO model file of our theoretical model with FeynRules~\cite{Christensen:2008py,Alloul:2013bka}, and then generate leading-order parton-level signal events with the MC-simulation tool \texttt{MadGraph5aMC$@$NLO}~\cite{Alwall:2014hca,Frederix:2018nkq}.
To account for next-to-leading-order (NLO) QCD corrections to the production cross sections, we rescale the leading-order results by a $K$-factor of 1.2 in our analysis.\footnote{Ref.~\cite{Ruiz:2015zca} reports NLO $K$-factors in the range $1.1$--$1.4$ for the pair production of type-III seesaw heavy leptons via Drell-Yan charged-current and neutral-current processes at the LHC. The production channels $H^{\pm} A$ and $HA$ considered in our work proceed through almost the same Drell-Yan charged-current and neutral-current mechanisms. We therefore adopt a representative $K$-factor of 1.2, following the prescription of Ref.~\cite{Ruiz:2015zca}.}
We expect that a full simulation including NLO QCD corrections would not substantially modify the viable parameter space.
The default cuts on the pseudorapidity and transverse momentum of the final-state jets and $b$-quarks are applied.
The produced LHE files~\cite{Alwall:2006yp} of the signal events are then fed to \texttt{Pythia8} for parton showering, hadronization, and completing the decay chains of various particles in the event records.
We then apply our recast and analysis codes to compute the cutflow efficiencies for all the parameter points scanned.
We are thus in position to compute the signal-event numbers $N_S$ with the following formula,
\bea
N_S=\mathcal{L}_\text{int.}\cdot \sigma \cdot \epsilon,
\eea
where $\mathcal{L}_\text{int.}=139\text{ fb}^{-1}$ or $3000\text{ fb}^{-1}$ is the integrated luminosity, $\sigma$ labels the signal-process cross section (summing up both the charged-current and the neutral current processes; see figure~\ref{figfeynman}), and $\epsilon$ denotes the cumulative cutflow efficiency obtained with our recast and analysis codes.

We note that for the original and the modified search analyses, we generate signal events at the center-of-mass energy $\sqrt{s}$ of 13 and 14 TeV, respectively.

\begin{figure}[t]
\centering
\includegraphics[width=0.5\textwidth]{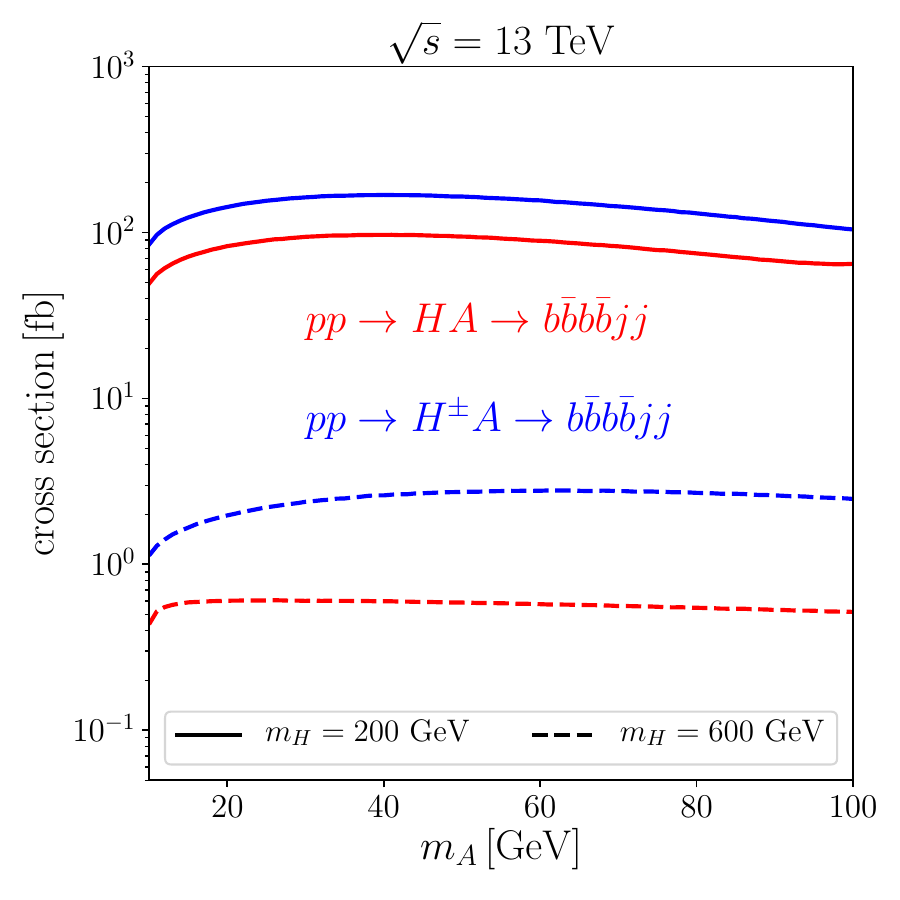}
\caption{Scattering cross sections of the two signal processes as functions of $m_A$, with $\sqrt{s}=13$~TeV. The red and blue lines are for the neutral- and charged-current signal processes, respectively, and the solid and dashed line styles are for $m_H=200$~GeV and $m_H=600$~GeV, respectively. We note that here we have enforced the intermediate $W$- and $Z$-bosons to each decay into two quark-jets including the $b$-quark.}
\label{figcs}
\end{figure}

In figure~\ref{figcs} we display the \texttt{MadGraph5}-computed scattering cross sections of our signal processes as functions of $m_A$, at the center-of-mass energy of 13 TeV.
Here, we have forced the $W$- and $Z$-bosons to decay into jets (including a $b$-jet).
Two benchmark values of $m_H$ are chosen: $m_H=200$~GeV (solid lines) and $m_H=600$~GeV (dashed lines).
As expected, for the case of a lighter $H$, the cross sections are larger; also, the charged-current process has larger cross sections than the neutral-current ones.
For the case of $\sqrt{s}=14$~TeV, the resulting cross sections are only slightly enhanced and are therefore not shown.

For the two analyses, we (expect to) have 0 background events.
Thus, for presenting numerical results, we will take 3 signal events as the exclusion limits at 95\% confidence level.

\section{Numerical results}
\label{sec:results}

In this section, we present and discuss numerical results.

\begin{table}[t]
\begin{center}
\resizebox{\textwidth}{!}{
\begin{tabular}{l|cc|cc|cc}
$m_A$~[GeV], $\tan{\beta}$, $c\tau_A$~[m]        & \multicolumn{2}{c|}{$15, 10^5, 4.2 \times 10^{-3}$} & \multicolumn{2}{c|}{$15, 10^6, 4.2 \times 10^{-1}$} & \multicolumn{2}{c}{$15, 10^7, 4.2 \times 10^{1}$} \\
\hline
Jet-$p_T$ threshold                            &    $5.4\times 10^{-3}$ & $3.3\times 10^{-2}$ &     $3.6\times 10^{-3}$  & $3.3\times 10^{-2}$ & $7.7\times 10^{-5}$   & $3.3\times 10^{-2}$  \\
Event has $\geq 1$ DV passing:                 &                        &                     &                          &                     &                       &                     \\
~~$R_{xy}, |z|<300$ mm               &          $5.3\times 10^{-3}$     & $3.3\times 10^{-2}$ &     $5.4\times 10^{-4}$  & $3.9\times 10^{-3}$ & $3.0\times 10^{-6}$   & $4.7\times 10^{-5}$  \\
~~$R_{xy}>4$ mm                      &          $5.2\times 10^{-3}$     & $3.2\times 10^{-2}$ &     $5.4\times 10^{-4}$  & $3.8\times 10^{-3}$ & $3.0\times 10^{-6}$   & $4.7\times 10^{-5}$  \\
~~$\geq 1$ trk.~with $|d_0| > 2$ mm &           $5.1\times 10^{-3}$     & $3.1\times 10^{-2}$ &     $5.3\times 10^{-4}$  & $3.8\times 10^{-3}$ & $3.0\times 10^{-6}$   & $4.7\times 10^{-5}$  \\
~~$n_{\text{sel.~dec.~prod.}}>5$     &          $5.1\times 10^{-3}$     & $3.0\times 10^{-2}$ &     $4.9\times 10^{-4}$  & $3.4\times 10^{-3}$ & $2.0\times 10^{-6}$   & $4.4\times 10^{-5}$  \\
~~$m_{\text{DV}}>10$ GeV             &          $8.9\times 10^{-4}$     & $4.9\times 10^{-3}$ &     $5.3\times 10^{-5}$  & $2.8\times 10^{-4}$ & $1.0\times 10^{-6}$   & $3.0\times 10^{-6}$  \\
\hline
Param.~Effi.                              &      $1.1\times 10^{-4}$    & $7.2\times 10^{-4}$ &     $8.0\times 10^{-7}$  & $2.0\times 10^{-5}$ & $6.0\times 10^{-9}$   & $8.5\times 10^{-8}$  \\
\hline
\hline

$m_A$~[GeV], $\tan{\beta}$, $c\tau_A$~[m]        & \multicolumn{2}{c|}{$30, 10^5, 2.1 \times 10^{-3}$} & \multicolumn{2}{c|}{$30, 10^6, 2.1 \times 10^{-1}$} & \multicolumn{2}{c}{$30, 10^7, 2.1 \times 10^{1}$} \\
\hline
Jet-$p_T$ threshold                           &     $5.4\times 10^{-3}$  & $3.0\times 10^{-2}$ &      $5.1\times 10^{-3}$    & $3.0\times 10^{-2}$ &   $2.7\times 10^{-4}$  & $3.0\times 10^{-2}$ \\
Event has $\geq 1$ DV passing:                &                          &                     &                             &                     &                        &                 \\
~~$R_{xy}, |z|<300$ mm               &     $5.4\times 10^{-3}$  & $3.0\times 10^{-2}$ &      $1.7\times 10^{-3}$    & $1.1\times 10^{-2}$ &   $1.6\times 10^{-5}$  & $1.5\times 10^{-4}$ \\
~~$R_{xy}>4$ mm                      &     $5.1\times 10^{-3}$  & $2.7\times 10^{-2}$ &      $1.7\times 10^{-3}$    & $1.0\times 10^{-2}$ &   $1.6\times 10^{-5}$  & $1.4\times 10^{-4}$ \\
~~$\geq 1$ trk.~with $|d_0| > 2$ mm&     $4.9\times 10^{-3}$  & $2.6\times 10^{-2}$ &      $1.7\times 10^{-3}$    & $1.0\times 10^{-2}$ &   $1.6\times 10^{-5}$  & $1.4\times 10^{-4}$ \\
~~$n_{\text{sel.~dec.~prod.}}>5$     &     $4.9\times 10^{-3}$  & $2.6\times 10^{-2}$ &      $1.6\times 10^{-3}$    & $9.6\times 10^{-3}$ &   $1.5\times 10^{-5}$  & $1.3\times 10^{-4}$ \\
~~$m_{\text{DV}}>10$ GeV            &     $4.3\times 10^{-3}$  & $2.3\times 10^{-2}$ &      $1.3\times 10^{-3}$    & $7.4\times 10^{-3}$ &   $9.0\times 10^{-6}$  & $9.7\times 10^{-5}$ \\
\hline
Param.~Effi.                              &     $1.7\times 10^{-3}$  & $1.1\times 10^{-2}$ &      $1.1\times 10^{-4}$     & $1.1\times 10^{-3}$ &   $2.3\times 10^{-7}$ & $9.7\times 10^{-6}$ \\

\hline
\hline
$m_A$~[GeV], $\tan{\beta}$, $c\tau_A$~[m]        & \multicolumn{2}{c|}{$60, 10^5, 1.1 \times 10^{-3}$} & \multicolumn{2}{c|}{$60, 10^6, 1.1 \times 10^{-1}$} & \multicolumn{2}{c}{$60, 10^7, 1.1 \times 10^{1}$}  \\
\hline
Jet-$p_T$ threshold                  &       $6.9\times 10^{-3}$ & $4.9\times 10^{-2}$ &      $6.8\times 10^{-3}$ & $4.9\times 10^{-2}$ &     $1.3\times 10^{-3}$    & $5.0\times 10^{-2}$  \\
Event has $\geq 1$ DV passing:       &                           &                     &                          &                     &                            &                         \\
~~$R_{xy}, |z|<300$ mm               &       $6.9\times 10^{-3}$ & $4.9\times 10^{-2}$ &      $4.8\times 10^{-3}$ & $3.5\times 10^{-2}$ &     $8.9\times 10^{-5}$    & $7.9\times 10^{-4}$  \\
~~$R_{xy}>4$ mm                      &       $4.4\times 10^{-3}$ & $2.8\times 10^{-2}$ &      $4.7\times 10^{-3}$ & $3.4\times 10^{-2}$ &     $8.9\times 10^{-5}$    & $7.8\times 10^{-4}$  \\
~~$\geq 1$ trk.~with $|d_0| > 2$ mm &       $4.2\times 10^{-3}$  & $2.7\times 10^{-2}$ &      $4.7\times 10^{-3}$ & $3.4\times 10^{-2}$ &     $8.9\times 10^{-5}$    & $7.8\times 10^{-4}$  \\
~~$n_{\text{sel.~dec.~prod.}}>5$     &       $4.2\times 10^{-3}$ & $2.7\times 10^{-2}$ &      $4.7\times 10^{-3}$ & $3.3\times 10^{-2}$ &     $8.8\times 10^{-5}$    & $7.5\times 10^{-4}$  \\
~~$m_{\text{DV}}>10$ GeV             &       $4.1\times 10^{-3}$ & $2.6\times 10^{-2}$ &      $4.6\times 10^{-3}$ & $3.3\times 10^{-2}$ &     $8.5\times 10^{-5}$    & $7.2\times 10^{-4}$  \\
\hline
Param.~Effi.                         &       $2.6\times 10^{-3}$ & $2.1\times 10^{-2}$ &      $1.3\times 10^{-3}$ & $1.1\times 10^{-2}$ &     $5.1\times 10^{-6}$    & $1.5\times 10^{-4}$  \\

\hline
\hline

$m_A$~[GeV], $\tan{\beta}$, $c\tau_A$~[m]  & \multicolumn{2}{c|}{$90, 10^5, 7.3 \times 10^{-4}$} & \multicolumn{2}{c|}{$90, 10^6, 7.3 \times 10^{-2}$} & \multicolumn{2}{c}{$90, 10^7, 7.3 \times 10^{0}$} \\
\hline
Jet-$p_T$ threshold                 &      $1.2\times 10^{-2}$     & $7.9\times 10^{-2}$ &    $1.2\times 10^{-2}$  & $7.9\times 10^{-2}$ & $4.0\times 10^{-3}$ & $7.9\times 10^{-2}$ \\
Event has $\geq 1$ DV passing:      &                              &                     &                         &                     &                     &                     \\
~~$R_{xy}, |z|<300$ mm              &      $1.2\times 10^{-2}$     & $7.9\times 10^{-2}$ &    $1.0\times 10^{-2}$  & $7.0\times 10^{-2}$ & $2.9\times 10^{-4}$ & $2.3\times 10^{-3}$ \\
~~$R_{xy}>4$ mm                     &      $3.9\times 10^{-3}$     & $2.0\times 10^{-2}$ &    $1.0\times 10^{-2}$  & $6.8\times 10^{-2}$ & $2.9\times 10^{-4}$ & $2.2\times 10^{-3}$ \\
~~$\geq 1$ trk.~with $|d_0| > 2$ mm &      $3.7\times 10^{-3}$     & $2.0\times 10^{-2}$ &    $1.0\times 10^{-2}$  & $6.8\times 10^{-2}$ & $2.9\times 10^{-4}$ & $2.2\times 10^{-3}$ \\
~~$n_{\text{sel.~dec.~prod.}}>5$    &      $3.7\times 10^{-3}$     & $2.0\times 10^{-2}$ &    $1.0\times 10^{-2}$  & $6.8\times 10^{-2}$ & $2.9\times 10^{-4}$ & $2.2\times 10^{-3}$ \\
~~$m_{\text{DV}}>10$ GeV            &      $3.7\times 10^{-3}$     & $1.9\times 10^{-2}$ &    $1.0\times 10^{-2}$  & $6.7\times 10^{-2}$ & $2.8\times 10^{-4}$ & $2.2\times 10^{-3}$ \\
\hline
Param.~Effi.                         &      $2.6\times 10^{-3}$     & $1.7\times 10^{-2}$ &    $4.4\times 10^{-3}$  & $3.4\times 10^{-2}$ & $1.7\times 10^{-5}$ & $5.9\times 10^{-4}$ \\
\hline
\end{tabular}
}
\caption{Cutflow efficiencies on one million MC-simulated events for $m_H=200$ GeV, considering $m_A=15, 30, 60$ and $90$ GeV, and $\tan{\beta}=10^{5}, 10^{6}$, and $10^{7}$. The corresponding $c\tau_A$ is computed with the \texttt{2HDMC} package. For each benchmark parameter chosen, there are two columns of efficiencies which correspond to the original and the modified analyses, respectively. The two analyses differ in their respective jet-$p_T$ thresholds and in the application of the parameterized efficiency functions; the modified analysis imposes looser jet-$p_T$ thresholds and does not apply the event-level parameterized efficiencies.}
\label{tab:cutflow_mH200}
\end{center}
\end{table}

\begin{table}[t]
\begin{center}
\resizebox{\textwidth}{!}{
\begin{tabular}{l|cc|cc|cc}
$m_A$~[GeV], $\tan{\beta}$, $c\tau_A$~[m]        & \multicolumn{2}{c|}{$15, 10^5, 4.2 \times 10^{-3}$} & \multicolumn{2}{c|}{$15, 10^6, 4.2 \times 10^{-1}$} & \multicolumn{2}{c}{$15, 10^7, 4.2 \times 10^{1}$}  \\
\hline
Jet-$p_T$ threshold                            &  $1.0\times 10^{-1}$ &  $3.3\times 10^{-1}$   &  $5.7\times 10^{-2}$   &  $3.3\times 10^{-1}$  &  $9.0\times 10^{-4}$  &  $3.3\times 10^{-1}$ \\
Event has $\geq 1$ DV passing:                 &                      &                        &                        &                       &                       &                      \\
~~$R_{xy}, |z|<300$ mm               &      $1.0\times 10^{-1}$       &  $3.2\times 10^{-1}$   &  $5.9\times 10^{-3}$   &  $2.1\times 10^{-2}$  &  $5.5\times 10^{-5}$  &  $2.1\times 10^{-4}$ \\
~~$R_{xy}>4$ mm                      &      $9.9\times 10^{-2}$       &  $3.2\times 10^{-1}$   &  $5.8\times 10^{-3}$   &  $2.0\times 10^{-2}$  &  $5.3\times 10^{-5}$  &  $2.1\times 10^{-4}$ \\
~~$\geq 1$ trk.~with $|d_0| > 2$ mm  &      $9.6\times 10^{-2}$       &  $3.1\times 10^{-1}$   &  $5.7\times 10^{-3}$   &  $2.0\times 10^{-2}$  &  $5.1\times 10^{-5}$  &  $2.0\times 10^{-4}$ \\
~~$n_{\text{sel.~dec.~prod.}}>5$     &      $9.6\times 10^{-2}$       &  $3.1\times 10^{-1}$   &  $5.6\times 10^{-3}$   &  $1.9\times 10^{-2}$  &  $5.0\times 10^{-5}$  &  $2.0\times 10^{-4}$ \\
~~$m_{\text{DV}}>10$ GeV             &      $1.8\times 10^{-2}$       &  $5.6\times 10^{-2}$   &  $6.8\times 10^{-4}$   &  $2.3\times 10^{-3}$  &  $6.0\times 10^{-6}$  &  $2.1\times 10^{-5}$ \\
\hline
Param.~Effi.                             &  $1.6\times 10^{-3}$       &  $6.7\times 10^{-3}$   &  $1.1\times 10^{-5}$   &  $1.4\times 10^{-4}$  &  $7.6\times 10^{-8}$  &  $1.4\times 10^{-6}$ \\
\hline
\hline
$m_A$~[GeV], $\tan{\beta}$, $c\tau_A$~[m]        & \multicolumn{2}{c|}{$30, 10^5, 2.1 \times 10^{-3}$} & \multicolumn{2}{c|}{$30, 10^6, 2.1 \times 10^{-1}$} & \multicolumn{2}{c}{$30, 10^7, 2.1 \times 10^{1}$} \\
\hline
Jet-$p_T$ threshold                            & $1.0\times 10^{-1}$ & $3.1\times 10^{-1}$ & $9.5\times 10^{-2}$ & $3.1\times 10^{-1}$ & $3.5\times 10^{-3}$ & $3.1\times 10^{-1}$  \\
Event has $\geq 1$ DV passing:                 &                     &                     &                     &                     &                     &                      \\
~~$R_{xy}, |z|<300$ mm               & $1.0\times 10^{-1}$ & $3.1\times 10^{-1}$ & $2.1\times 10^{-2}$ & $6.9\times 10^{-2}$ & $2.2\times 10^{-4}$ & $8.6\times 10^{-4}$  \\
~~$R_{xy}>4$ mm                      & $9.8\times 10^{-2}$ & $3.0\times 10^{-1}$ & $2.1\times 10^{-2}$ & $6.8\times 10^{-2}$ & $2.2\times 10^{-4}$ & $8.4\times 10^{-4}$  \\
~~$\geq 1$ trk.~with $|d_0| > 2$ mm  & $9.5\times 10^{-2}$ & $2.9\times 10^{-1}$ & $2.1\times 10^{-2}$ & $6.7\times 10^{-2}$ & $2.2\times 10^{-4}$ & $8.4\times 10^{-4}$  \\
~~$n_{\text{sel.~dec.~prod.}}>5$     & $9.5\times 10^{-2}$ & $2.9\times 10^{-1}$ & $2.1\times 10^{-2}$ & $6.7\times 10^{-2}$ & $2.2\times 10^{-4}$ & $8.2\times 10^{-4}$  \\
~~$m_{\text{DV}}>10$ GeV             & $8.7\times 10^{-2}$ & $2.7\times 10^{-1}$ & $1.7\times 10^{-2}$ & $5.5\times 10^{-2}$ & $1.9\times 10^{-4}$ & $6.6\times 10^{-4}$  \\
\hline
Param.~Effi.                             & $3.0\times 10^{-2}$ & $1.2\times 10^{-1}$ & $1.2\times 10^{-3}$ & $8.8\times 10^{-3}$ & $5.1\times 10^{-6}$ & $9.7\times 10^{-5}$  \\
\hline

\hline       
$m_A$~[GeV], $\tan{\beta}$, $c\tau_A$~[m]        & \multicolumn{2}{c|}{$60, 10^5, 1.1 \times 10^{-3}$} & \multicolumn{2}{c|}{$60, 10^6, 1.1 \times 10^{-1}$} & \multicolumn{2}{c}{$60, 10^7, 1.1 \times 10^{1}$}  \\
\hline
Jet-$p_T$ threshold                            & $1.0\times 10^{-1}$ & $3.6\times 10^{-1}$ & $1.0\times 10^{-1}$ & $3.6\times 10^{-1}$ & $1.2\times 10^{-2}$ & $3.6\times 10^{-1}$  \\
Event has $\geq 1$ DV passing:                 &                     &                     &                     &                     &                     &                     \\
~~$R_{xy}, |z|<300$ mm               &           $1.0\times 10^{-1}$ & $3.6\times 10^{-1}$ & $5.6\times 10^{-2}$ & $2.1\times 10^{-1}$ & $8.7\times 10^{-4}$ & $3.5\times 10^{-3}$  \\
~~$R_{xy}>4$ mm                      &           $7.6\times 10^{-2}$ & $2.6\times 10^{-1}$ & $5.5\times 10^{-2}$ & $2.1\times 10^{-1}$ & $8.6\times 10^{-4}$ & $3.4\times 10^{-3}$  \\
~~$\geq 1$ trk.~with $|d_0| > 2$ mm &            $7.2\times 10^{-2}$ & $2.5\times 10^{-1}$ & $5.5\times 10^{-2}$ & $2.1\times 10^{-1}$ & $8.6\times 10^{-4}$ & $3.4\times 10^{-3}$  \\
~~$n_{\text{sel.~dec.~prod.}}>5$     &           $7.2\times 10^{-2}$ & $2.5\times 10^{-1}$ & $5.5\times 10^{-2}$ & $2.1\times 10^{-1}$ & $8.6\times 10^{-4}$ & $3.4\times 10^{-3}$  \\
~~$m_{\text{DV}}>10$ GeV             &           $7.1\times 10^{-2}$ & $2.4\times 10^{-1}$ & $5.4\times 10^{-2}$ & $2.0\times 10^{-1}$ & $8.2\times 10^{-4}$ & $3.3\times 10^{-3}$  \\
\hline
Param.~Effi.                             &       $3.8\times 10^{-2}$ & $1.8\times 10^{-1}$ & $1.2\times 10^{-2}$ & $6.9\times 10^{-2}$ & $4.2\times 10^{-5}$ & $8.5\times 10^{-4}$  \\
\hline
\hline       
$m_A$~[GeV], $\tan{\beta}$, $c\tau_A$~[m]        & \multicolumn{2}{c|}{$90, 10^5, 7.3 \times 10^{-4}$} & \multicolumn{2}{c|}{$90, 10^6, 7.3 \times 10^{-2}$} & \multicolumn{2}{c}{$90, 10^7, 7.3 \times 10^{0}$} \\
\hline
Jet-$p_T$ threshold                            & $1.6\times 10^{-1}$ & $4.9\times 10^{-1}$ & $1.6\times 10^{-1}$ & $4.9\times 10^{-1}$ & $4.0\times 10^{-2}$ & $4.9\times 10^{-1}$ \\
Event has $\geq 1$ DV passing:                 &                     &                     &                     &                     &                     &                 \\
~~$R_{xy}, |z|<300$ mm               & $1.6\times 10^{-1}$ & $4.9\times 10^{-1}$ & $1.3\times 10^{-1}$ & $4.0\times 10^{-1}$ & $3.0\times 10^{-3}$ & $1.0\times 10^{-2}$ \\
~~$R_{xy}>4$ mm                      & $6.9\times 10^{-2}$ & $1.9\times 10^{-1}$ & $1.3\times 10^{-1}$ & $4.0\times 10^{-1}$ & $3.0\times 10^{-3}$ & $1.0\times 10^{-2}$ \\
~~$\geq 1$ trk.~with $|d_0| > 2$ mm & $6.5\times 10^{-2}$ & $1.8\times 10^{-1}$ & $1.3\times 10^{-1}$ & $3.9\times 10^{-1}$ & $3.0\times 10^{-3}$ & $1.0\times 10^{-2}$ \\
~~$n_{\text{sel.~dec.~prod.}}>5$      & $6.5\times 10^{-2}$ & $1.8\times 10^{-1}$ & $1.3\times 10^{-1}$ & $3.9\times 10^{-1}$ & $3.0\times 10^{-3}$ & $9.9\times 10^{-3}$ \\
~~$m_{\text{DV}}>10$ GeV             & $6.4\times 10^{-2}$ & $1.8\times 10^{-1}$ & $1.3\times 10^{-1}$ & $3.9\times 10^{-1}$ & $2.9\times 10^{-3}$ & $9.8\times 10^{-3}$ \\
\hline
Param.~Effi.                            & $3.8\times 10^{-2}$ & $1.5\times 10^{-1}$ & $4.7\times 10^{-2}$ & $1.9\times 10^{-1}$ & $1.8\times 10^{-4}$ & $3.0\times 10^{-3}$ \\
\hline
\end{tabular}
}
\caption{The same as table~\ref{tab:cutflow_mH200} but for $m_H=600$ GeV.}
\label{tab:cutflow_mH600}
\end{center}
\end{table}

We first list in table~\ref{tab:cutflow_mH200} and table~\ref{tab:cutflow_mH600} cutflow efficiencies based on one million signal events simulated at each representative parameter point selected, for $m_H=200$~GeV and $m_H=600$~GeV, respectively.
We have chosen combinations of $m_A=15, 30, 60, \text{ and }90$ GeV and $\tan\beta=10^5, 10^6,\text{ and }10^7$, for displaying these cutflow efficiencies.
In particular, for each parameter combination, we give also the corresponding proper decay length $c\tau_A$ computed with the \texttt{2HDMC} package, and list the cutflow efficiencies with both the original and the modified analyses in two columns for the purpose of easier comparison.
Indeed, as expected, the latter shows that the modified analysis is predicted to have higher final cutflow efficiencies in all parameter combinations.

Among the selected parameter combinations, for a fixed value of $m_A$, increasing $\tan\beta$ leads to fewer events passing the fiducial-volume requirement.
Also, for a heavier $A$, the relative probabilities of passing the requirements on the number of selected decay products and the DV invariant mass increase; the increased number of selected decay products is mainly due to the enlarged phase space leading to enhanced production of more sufficiently boosted tracks in $A$-decays.

Furthermore, comparing the two tables, we find that in the case of a larger value of $m_H$, the cutflow efficiencies are higher by roughly one order of magnitude, mainly arising from the higher efficiencies to pass the jet-$p_T$-threshold requirements.
This is primarily because the jets from decays of $H$ and $H^\pm$ tend to be harder and more abundant for larger values of $m_H=m_{H^\pm}$.

In these two tables, we observe that the largest and smallest final cutflow efficiencies are, respectively, in the order of $10^{-9}$ and $10^{-1}$.

\begin{figure}[tb]
\centering
\includegraphics[width=0.9\textwidth]{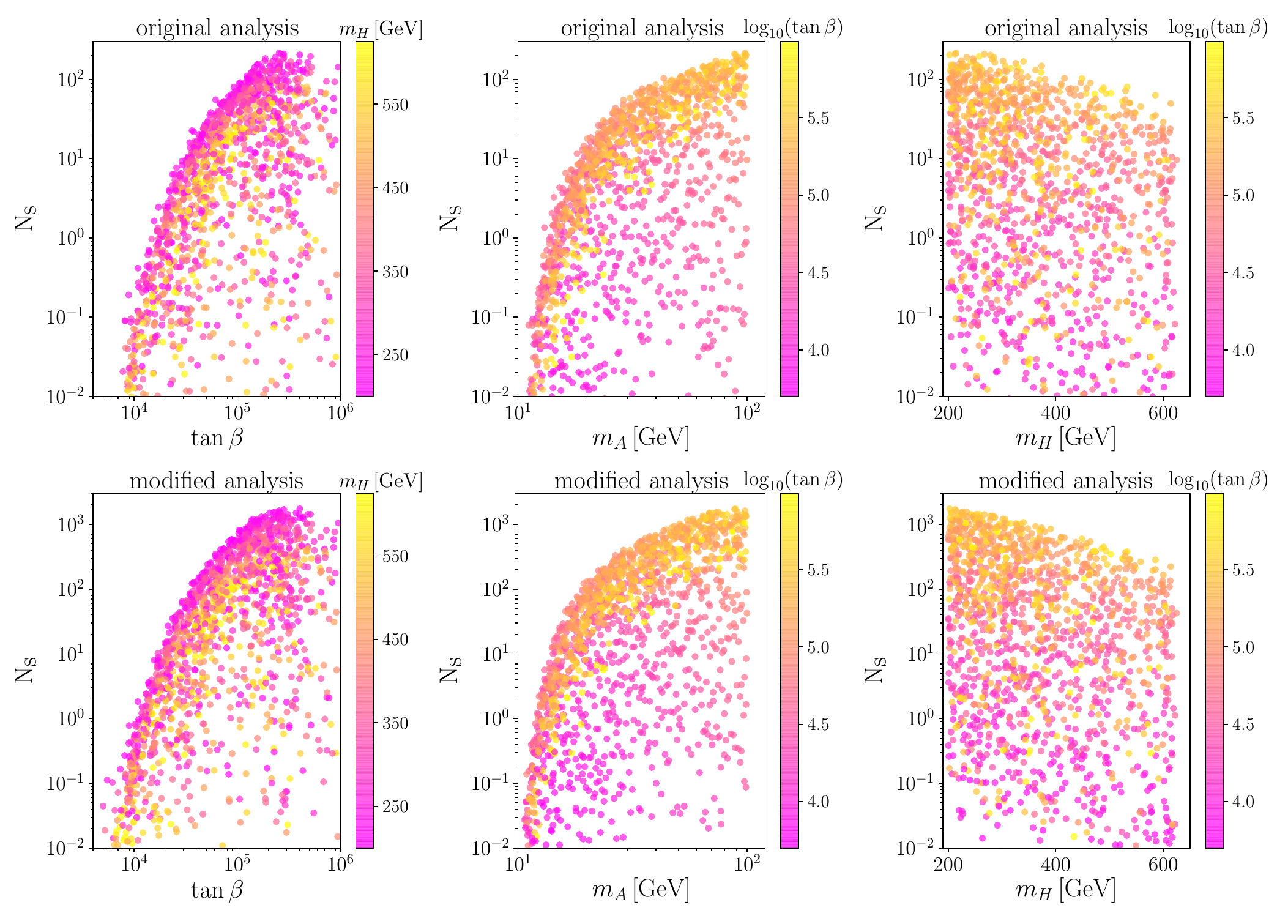}
\caption{Signal-event numbers $N_S$ with an integrated luminosity of $\mathcal{L}_{\text{int.}}=139\text{ fb}^{-1}$, for the original (upper panels) and the modified (lower panels) analyses. The three columns are for $N_S$ vs.~$\tan{\beta}$ at different $m_H$ levels, $N_S$ vs.~$m_A$ at different $\tan{\beta}$ levels, and $N_S$ vs.~$m_H$ at different $\tan{\beta}$ levels, respectively.}
\label{fignum}
\end{figure}

We proceed to display in figure~\ref{fignum} multiple plots of signal-event numbers as functions of $\tan\beta$, $m_A$, and $m_H$, for a fixed integrated luminosity of 139 fb$^{-1}$.
The two rows of plots in figure~\ref{fignum} are for the original analysis and the modified analysis, respectively.
We note that on each displayed point in these plots, only two parameters among $(m_A, m_H, \tan\beta)$ are fixed, while the remaining one can be of any value as long as both the theoretical and experimental constraints are obeyed.
The behavior of the displayed points on each plot can be easily explained qualitatively.
For instance, the plots in the first column indicate that $N_S$ initially rises and then falls as $\tan\beta$ increases.
Since increasing $\tan\beta$ renders $A$ longer lived, this behavior corresponds to the transition of the $A$ particle from being promptly decaying, to being displaced decaying inside the inner detector, and finally to being decaying more often behind the inner detector.
The plots in the second column show enhanced signal-event rates in general, for larger $m_A$.
This is because a heavier $A$ results in higher rates of passing the selection criteria on the jet-$p_T$ thresholds, the number of selected decay products, and the invariant mass of the DV.
Finally, in the last column's plots we observe declining number of signal events for a heavier $H$.
While a heavier $H$ leads to larger efficiencies to pass the event-selection criteria, it also leads to reduction in the signal-event cross sections (see figure~\ref{figcs}). The combination of both effects, in the end, leads to a decrease in $N_S$.

\begin{figure}[tb]
\centering
\includegraphics[width=0.9\textwidth]{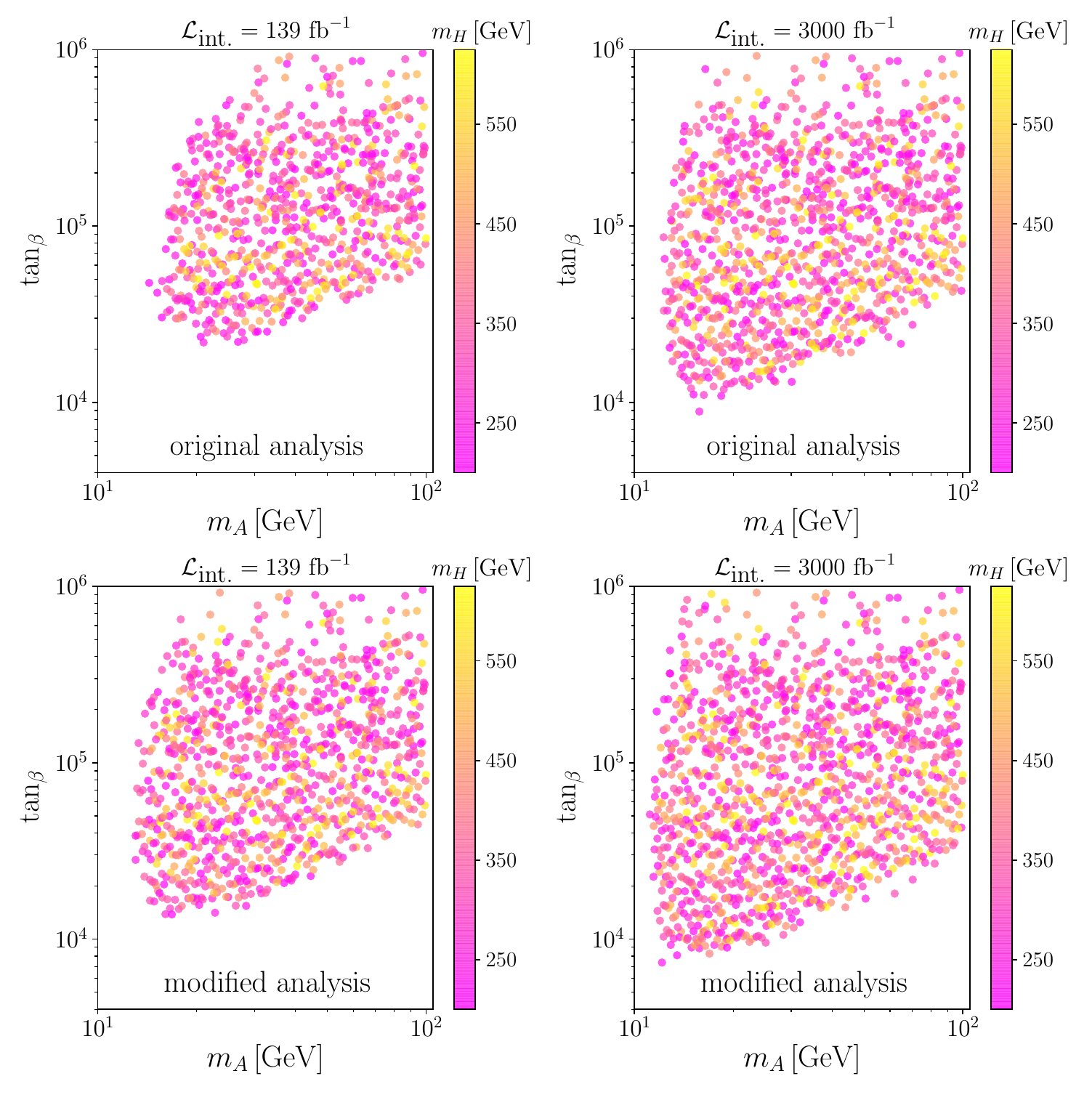}
\caption{Parameter points where at least 3 signal events are predicted, shown in the $\tan\beta$ vs.~$m_A$ plane for various levels of $m_H$. The two rows correspond to the original and the modified analyses, respectively. The left (right) panels are for an integrated luminosity of 139 fb$^{-1}$ (3000 fb$^{-1}$).}
\label{fignum3}
\end{figure}

We display in figure~\ref{fignum3} a set of plots of parameter points for which at least 3 signal events are predicted by our computation, all shown in the $\tan\beta$ vs.~$m_A$ plane, for different values of $m_H$.
The upper and lower rows are obtained with the original and the modified analyses, respectively, and the two columns are for integrated luminosities of 139 fb$^{-1}$ and 3000 fb$^{-1}$, respectively.
On all the shown parameter points, the current theoretical and experimental bounds are satisfied.

We find that the modified analysis can probe both larger and smaller values of $\tan\beta$ on the two ends than the original analysis.
Also, a larger integrated luminosity allows to test broader regions in the parameter space.
The sensitivities are all cut off at positions just above $m_A=10$ GeV, as a result of the $m_{\text{DV}}>10$~GeV requirement.
For too small values of $\tan\beta$, the $A$ particle is too promptly decaying, failing to decay at $r>4$ mm.

\begin{figure}[tb]
\centering
\includegraphics[width=0.495\textwidth]{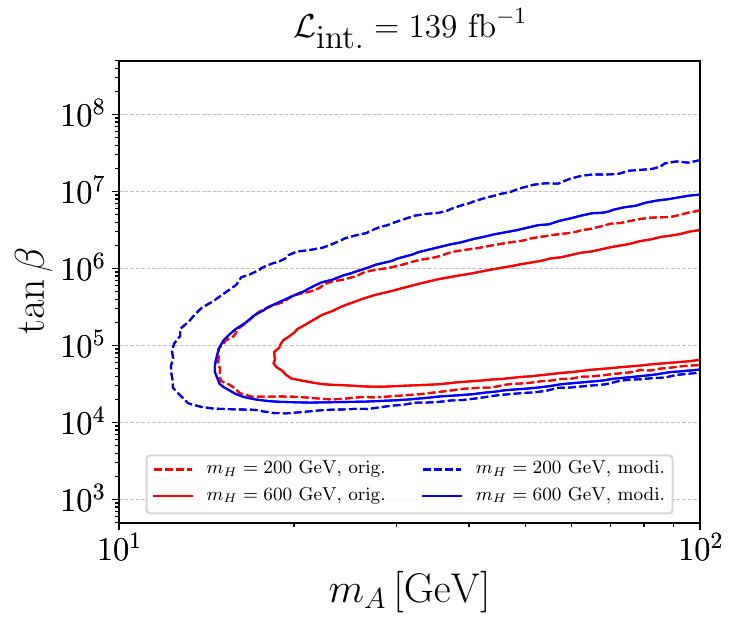}
\includegraphics[width=0.495\textwidth]{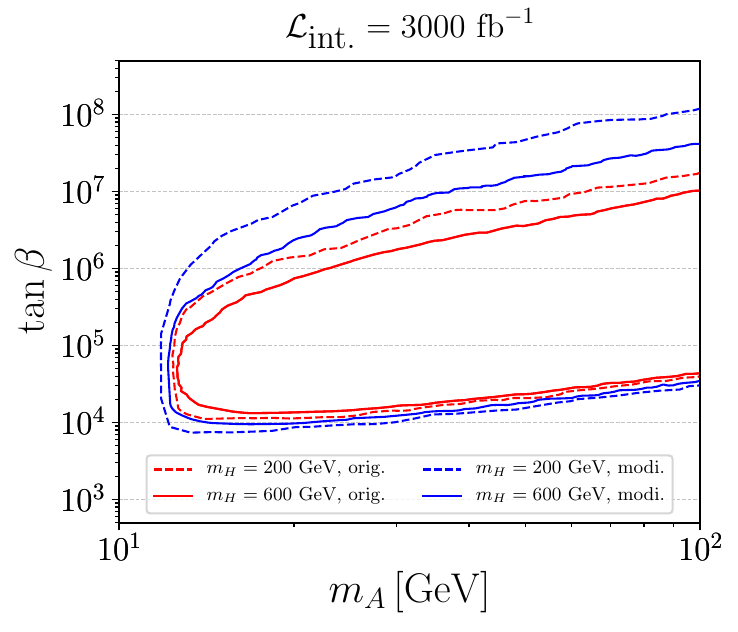}
\caption{$95\%$ C.L.~sensitivity reach to $\tan\beta$ with respect to $m_A$, for $m_H=200$~GeV (dashed lines) and $m_H=600$~GeV (solid lines). The left (right) panel is for $\mathcal{L}_{\text{int.}}=139$~fb$^{-1}$ ($\mathcal{L}_{\text{int.}}=3000$~fb$^{-1}$), and the red (blue) isocurves correspond to the original (modified) analysis.}
\label{fig:sensitivity}
\end{figure}

Following the discussion above, we present two additional sensitivity plots in figure~\ref{fig:sensitivity}, corresponding to integrated luminosities of 139 fb$^{-1}$ and 3000 fb$^{-1}$, respectively.
In each plot, we choose two benchmark values of $m_H$ at 200 GeV (dashed) and 600 GeV (solid), and present contour curves of 3 signal events\footnote{If, instead of vanishing background levels, we have 1 or 10 background events, the sensitivity results are expected to weaken by an almost negligible extent.} with the original (red) and the modified (blue) analyses, corresponding to the 95\% C.L.~exclusion limits in the condition of vanishing background.
The results show clearly that the sensitivities shrink for a heavier $H$, and that the modified analysis is expected to probe larger parameter regions than the original one.
With an integrated luminosity of 3000~fb$^{-1}$, the modified analysis with $m_H=200$~GeV can probe $\tan\beta\lesssim 10^{8}$ for $m_A\lesssim 100$~GeV (for $\tan\beta\sim10^8$ and $m_A\sim 100$~GeV, $c\tau_A\sim 650$~m).\footnote{We note that while in section~\ref{sec:currentlimits} we state that we will restrict our analysis to the region $\tan\beta\lesssim 10^6$, here we extend the study to values of $\tan\beta$ up to about $10^8$. Such large values of $\tan\beta$ require excessive fine-tuning, and checking the theoretical constraints becomes challenging because of numerical precision limitations.}
For even larger values of $\tan\beta$, $A$ becomes too long-lived to decay before leaving the inner detector.

\section{Conclusions}
\label{sec:conclu}

In this work, we have shown that for sufficiently large values of $\tan\beta$ in the type-I 2HDM, the pseudoscalar $A$ can acquire a macroscopic lifetime, leading to distinctive signatures of DVs plus jets inside the ATLAS and CMS inner detectors.
The dominant production of $A$ proceeds via electroweak processes, $pp \rightarrow W^{\pm*}/Z^* \to H^\pm/H\,A$, followed by the subsequent decays $H^\pm/H \to W^{\pm}/Z\,A$.
In addition, we have confined ourselves to the signal decays of the SM $W$- and $Z$-bosons into quark-jets including the $b$-quark, and that of $A$ into a pair of $b$-quarks.

Through MC simulations, we have analyzed the signal events characterized by DVs associated with multiple jets.
We have evaluated the discovery prospects of $A$ with an ``original analysis'' and a ``modified analysis'' separately, based on the recast of a recent ATLAS search for DVs plus jets and a past 8-TeV ATLAS search for DVs, considering both an integrated luminosity of $139\text{~fb}^{-1}$ and $3000\text{~fb}^{-1}$.
The numerical results of our study demonstrate that searches for DVs plus multiple jets provide a promising discovery avenue for a long-lived pseudoscalar $A$ predicted in the type-I 2HDM framework.
We find that a large region of parameter space with $10\text{~GeV}<m_A<100\text{~GeV}$ has already been excluded by the LHC Run-2 data, and future upgrades in the HL-LHC era can effectively allow to test broader regions for the same range of $m_A$, especially with the proposed modified analysis.

\acknowledgments
L.~W.~acknowledges support by the Projects No.~ZR2024MA001 and No.~ZR2023MA038 supported by Shandong Provincial Natural Science Foundation and by the National Natural Science Foundation of China under grants No.11975013.
Z.~S.~W.~acknowledges support by the National Natural Science Foundation of China under Grant Nos.~12475106 and 12505120 and the Fundamental Research Funds for the Central Universities under Grant No.~JZ2025HGTG0252.

\bibliographystyle{JHEP}
\bibliography{refs.bib}

\end{document}